\documentclass[preprint2,twocolumn,times,tighten]{aastex631}
\usepackage{graphicx}	
\usepackage{amsmath}	
\usepackage{amssymb}	
\usepackage{multirow}
\usepackage{booktabs}
\usepackage[dvipsnames]{xcolor}
\let\tablenum\relax
\usepackage{siunitx}
\usepackage[normalem]{ulem}

\begin{document}
\title[Galactic Dust Polarization in Turbulent Multiphase ISM: On the Origin of the $EE/BB$ Asymmetry]{Galactic Dust Polarization in Turbulent Multiphase ISM: On the Origin of the $EE/BB$ Asymmetry}

\email{yuehu@ias.edu; *NASA Hubble Fellow}

\author[0000-0002-8455-0805]{Yue Hu*}
\affiliation{Institute for Advanced Study, 1 Einstein Drive, Princeton, NJ 08540, USA}
\affiliation{Cahill Center for Astronomy and Astrophysics, California Institute of Technology, Pasadena, CA, USA}

\author[0000-0001-9654-8051]{Bao Truong}
\affiliation{Korea Astronomy and Space Science Institute, Daejeon 34055, Republic of Korea}
\affiliation{Korea University of Science and Technology, 217 Gajeong-ro, Yuseong-gu, Daejeon, 34113, Republic of Korea}

\author[0000-0003-2017-0982]{Thiem Hoang}
\affiliation{Korea Astronomy and Space Science Institute, Daejeon 34055, Republic of Korea}
\affiliation{Korea University of Science and Technology, 217 Gajeong-ro, Yuseong-gu, Daejeon, 34113, Republic of Korea}

\author[0000-0002-6488-8227]{Le Ngoc Tram}
\affiliation{Leiden Observatory, Leiden University, PO Box 9513, 2300 RA Leiden, The Netherlands}



\begin{abstract}
Polarized thermal emission from Galactic dust is the dominant foreground for CMB polarization measurements at high frequencies, with its statistical properties shaped by the interplay between turbulence and magnetic fields in the multiphase interstellar medium (ISM). Variations in turbulence regime and density–magnetic-field alignment across the warm (WNM), unstable (UNM), and cold (CNM) neutral media should imprint distinct signatures on the power spectra and $EE/BB$ power ratio, yet the relative contributions of these phases remain poorly constrained. Using high-resolution 3D magnetohydrodynamic simulations of a turbulent multiphase ISM coupled with synthetic dust polarization maps, we quantify phase-dependent turbulence, anisotropy, and alignment properties. We find that the trans-Alfv\'enic and transonic WNM and UNM are strongly anisotropic, exhibiting tight alignment of density and velocity structures with the local magnetic field. In contrast, the super-Alfv\'enic and supersonic CNM displays reduced anisotropy and weak alignment. These dynamical differences are reflected in the statistical scaling of fluctuations: the square root of the second-order velocity structure function exhibits a slope near $1/3$ in the WNM, near $1/2$ in the CNM, and intermediate in the UNM. Comparing our synthetic polarization power spectra with \textit{Planck} measurements, we find that polarization from UNM dust yields spectral slopes closest to the \textit{Planck}-inferred values, whereas WNM and CNM dust produce steeper and shallower spectra, respectively. The WNM yields $EE/BB>2$, the UNM gives $EE/BB\sim2$, and the CNM yields $EE/BB\approx1$. These results suggest that UNM dust may be an important contributor to the polarized foreground under typical high-latitude ISM conditions. We present predictions at 150 GHz to inform foreground modeling and separation.
\end{abstract}



\section{Introduction} \label{sec:intro}
The interstellar medium (ISM) is a highly dynamic, multiphase environment governed by the complex interplay of gas, dust, magnetic fields, and turbulence across a vast range of spatial  scales \citep{1977ApJ...218..148M,1995ApJ...443..152W,1999MNRAS.302..417S,2000ApJ...540..271V,2000A&A...354..247L,2010ApJ...710..853C,2011piim.book.....D,2011Sci...334..955L,2012ARA&A..50...29C,2012ARA&A..50..491P,Gray2017,2019NatAs...3..776H,2022ApJ...941..133H,2022MNRAS.510.4952L,2022ApJ...934....7H,2024A&A...691A.303T,2025ApJ...990..165V,2025ApJ...988..188H}. This multiphase structure dictates fundamental astrophysical processes, from cosmic ray transport and acceleration \citep{1966ApJ...146..480J,1969ApJ...155..777J,2002ApJ...578L.117Q,2007ARNPS..57..285S,2009Natur.462..770V,2020ApJ...894...63X,2021ApJ...923...53L,2023MNRAS.520.5126K,2025ApJ...994..142H,2025arXiv250907104H,2025FrASS..1111076M} to star formation \citep{2007ARA&A..45..565M,2009MNRAS.398.1082B,2012ApJ...761..156F,2012ARA&A..50...29C,2019NatAs...3..776H,2020ApJ...901..162H,2022MNRAS.515.4929G,2025arXiv251012203F,vazquez2025interstellar,2026RAA....26c5001A} and the broader evolution of galaxies \citep{2013MNRAS.433.1970F,2017ApJ...837..150S,2017MNRAS.466...88S,2018MNRAS.477.2716K,2018MNRAS.478.1694M,2023MNRAS.526..224W,2025ApJ...983...32H}. Within this environment, the alignment of asymmetric dust grains with the local magnetic field produces polarized thermal emission, serving as a critical observational probe of both ISM magnetism and fundamental dust physics \citep{2007MNRAS.378..910L,2008MNRAS.388..117H,2011piim.book.....D,2015ARA&A..53..501A,2021ApJ...909...94D,2022ApJ...926...90D,2025ApJ...994..115H}.

However, this same polarized dust emission constitutes the dominant Galactic foreground for cosmic microwave background (CMB) polarization measurements \citep{2014PTEP.2014fB109I,2014PhRvL.112x1101B,2016A&A...594A..10P,2016A&A...586A.133P,2019JCAP...02..056A,2019A&A...632A..17B,2020A&A...641A..11P,2022ApJ...926...54A,2026ApJ...999...89A}. At frequencies above 70 GHz, thermal dust polarization severely obscures the faint primordial $B$-mode signal—a theorized signature of gravitational waves generated during cosmic inflation \citep{2001PhRvD..64j3001Z,2014A&A...571A..11P,2014A&A...566A..55P,2015A&A...576A.105P,2020A&A...641A..11P}. A robust detection of these $B$-modes would offer unprecedented insights into the physics of the early Universe, making the accurate characterization and removal of the dust foreground a paramount challenge in modern cosmology \citep{2016ARA&A..54..227K,2017A&A...603A..62V,2019JCAP...02..056A}.

Effective foreground separation requires a physically grounded understanding of how magnetized interstellar structures shape the complex morphology and statistical properties of the polarized sky \citep{2018PhRvL.121b1104K,2019ApJ...870...87B,2019ApJ...880..106K,2020ApJ...899...31H,2023ARA&A..61...19M,2024ApJ...972...26S}. Modeling this emission from first principles is notoriously difficult due to the highly compressible, nonlinear, and turbulent nature of the multiphase ISM \citep{1999MNRAS.302..417S,2000ApJ...540..271V,2009ARA&A..47...27K,2011piim.book.....D,2018JCAP...08..049B,vazquez2025interstellar}. Observational and phenomenological studies suggest that the preferential alignment of filamentary dust structures with the magnetic field generates the observed $EE/BB$ asymmetry, where $E$-mode power strictly exceeds $B$-mode power \citep{2001PhRvD..64j3001Z,2017ApJ...839...91C,2017A&A...601A..71G,2020ApJ...899...31H,2022A&A...663A.175K}. Crucially, because this alignment is heavily regulated by magnetohydrodynamic (MHD) turbulence \citep{2018MNRAS.478..530K}, the resulting polarization signatures should vary significantly across distinct ISM phases. Although prior studies have established the general picture of the $EE/BB$ asymmetry \citep{2017A&A...601A..71G,2018PhRvL.121b1104K, 2019ApJ...880..106K,2025PhRvD.112j1302H}, a statistical characterization of the turbulence properties within each phase—and their quantitative relative contributions to the integrated global polarization signal—remains unexplored. Because the mass and volume fractions of the warm neutral medium (WNM), unstable neutral medium (UNM), and cold neutral medium (CNM) fluctuate significantly across different lines of sight (LOS) in the Galactic sky \citep{2009ARA&A..47...27K,2023ARA&A..61...19M}, disentangling these phase-specific signatures is essential for accurately characterizing and removing dust foregrounds in next-generation CMB experiments.

Fully resolving the complex dynamics of the interstellar medium—from the large-scale turbulence injection down to the typical dense CNM scale of 0.1 -- 10~pc—demands very high numerical resolution. In this study, we investigate foreground dust polarization using $2048^3$-grid 3D magnetohydrodynamic (MHD) simulations of the turbulent multiphase ISM. By capturing spatial scales from 100~pc down to 0.05~pc, our simulations self-consistently evolve gas, turbulence, and magnetic fields across multiple thermal phases. Achieving this extreme resolution guarantees a sufficiently broad inertial range, which is strictly required to simultaneously resolve the large-scale turbulent driving and the fine, small-scale structures that govern the observed polarization. This allows us to explicitly characterize the physical properties of the WNM, UNM, and CNM, and evaluate their alignment with the local magnetic field. While pioneering previous studies \citep[e.g.,][]{2018PhRvL.121b1104K} have conducted a parameter sweep of magnetization at $512^3$ resolution, our work focuses on the turbulence properties and dust polarization under typical diffuse ISM conditions, ensuring that the WNM, UNM, and critically, the CNM structures, are fully and physically resolved.

Previous synthetic polarization studies based on MHD turbulence have typically relied on the assumption of uniform grain alignment and have not investigated the distinct contributions of different ISM phases \citep{2015A&A...576A.105P,2018PhRvL.121b1104K,2025PhRvD.112j1302H}. In this work, we generate synthetic dust polarization maps by post-processing our simulations with an upgraded version of the \textsc{Polaris} code, which incorporates state-of-the-art grain alignment physics \citep{Reissl2016,Giang2023}. We also separate the contribution from WNM, UNM, and CNM. This allows us to quantify how the distinct phases of the ISM modulate key polarization properties, including the $E$- and $B$-mode power spectra and the $EE/BB$ asymmetry. Such analysis is critical for refining foreground separation strategies in CMB experiments. Furthermore, we go beyond comparisons with \textit{Planck} observations at 353 GHz to present quantitative predictions for future CMB surveys at 150 GHz, establishing a physically motivated framework for the improved removal of polarized foregrounds in the search for primordial $B$-modes.

This paper is organized as follows. \S~\ref{sec:method} describes the 3D multi-phase ISM simulations employed in this study and outlines the generation of synthetic dust polarization and the associated analysis methods. In \S~\ref{sec:results}, we present statistical analyses of the velocity, density, and magnetic field, followed by the statistics of dust polarization and the $E$- and $B$-mode power spectra. We compare our results with earlier work in \S~\ref{sec:discussion} and conclude in \S~\ref{sec:conclusion} with a summary of the main results.

\begin{figure*}[p]
\centering
\includegraphics[width=0.9\linewidth]{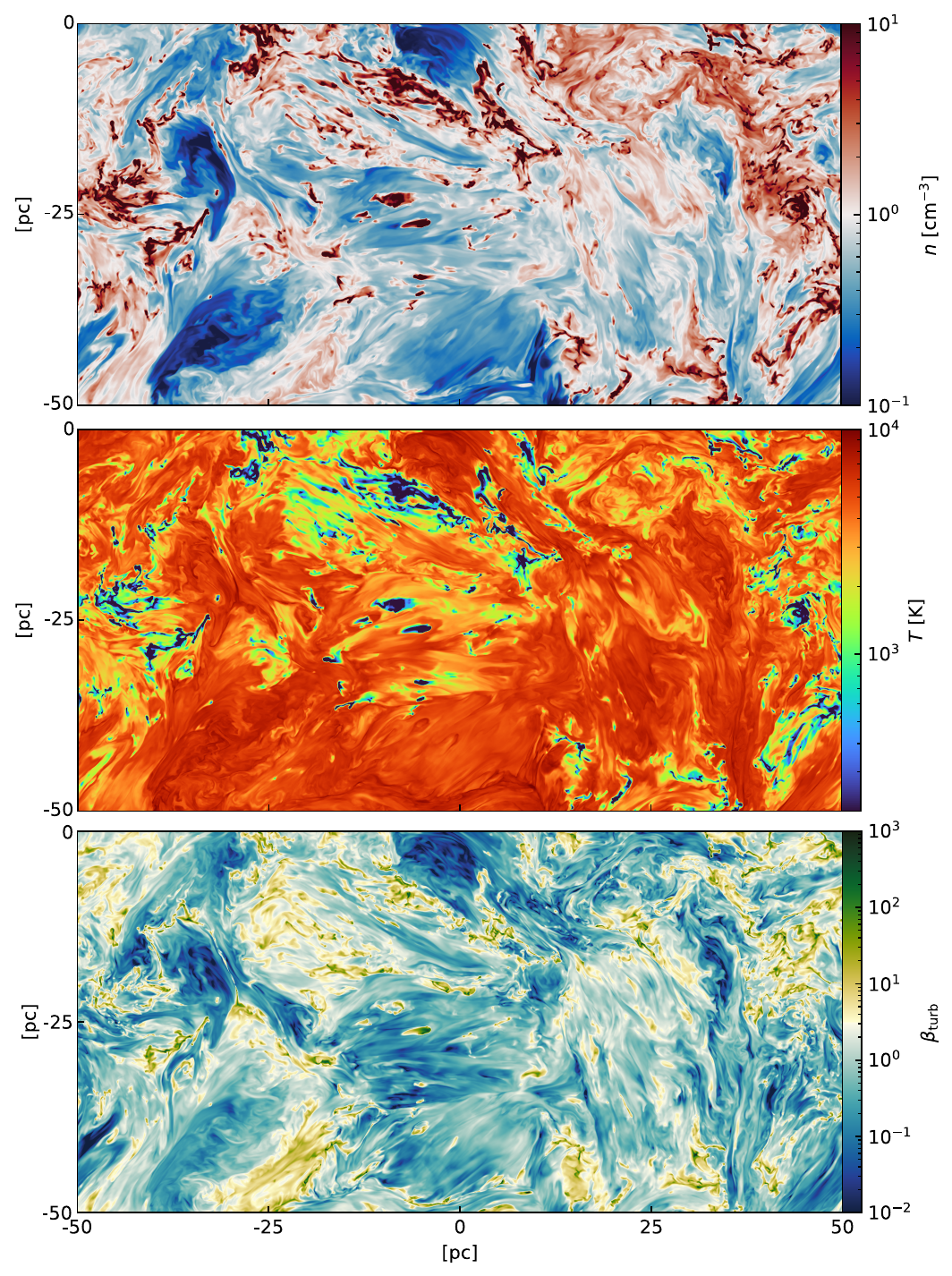}
\caption{Distributions of number density $n$, temperature $T$, and turbulent plasma beta $\beta_{\rm tur}$ in the multiphase ISM simulation. Shown is a two-dimensional slice from the $2048^3$ simulation, displaying the number density (top), temperature (middle), and turbulent plasma beta $\beta_{\rm tur} = \rho v^2 / (B^2 / 8\pi)$ (bottom), where $\rho$ is the gas mass density and $v$ is the local turbulent velocity. For visualization purposes, only half of each slice is shown. The cold neutral medium, characterized by low temperature, is spatially correlated with the densest regions.}
\label{fig:nTv_map}
\end{figure*}

\section{Methodology} 
\label{sec:method}
\subsection{MHD simulations of multi-phase ISM} 
The 3D multi-phase ISM simulations analyzed in this study were generated using the AthenaK code \citep{2024arXiv240916053S,2025ApJ...986...62H}. These simulations solve the ideal MHD equations with periodic boundary conditions, given by:
\begin{equation} 
\label{eq.mhd}
\begin{aligned}
&\frac{\partial\rho}{\partial t} +\nabla\cdot(\rho\boldsymbol{v})=0,\\
&\frac{\partial(\rho\boldsymbol{v})}{\partial t}+\nabla\cdot\left[\rho\boldsymbol{v}\boldsymbol{v}^T+\left(P+\frac{B^2}{8\pi}\right)\boldsymbol{I}-\frac{\boldsymbol{B}\boldsymbol{B}^T}{4\pi}\right] = \boldsymbol{f},\\
&\frac{\partial\boldsymbol{B}}{\partial t} - \nabla\times(\boldsymbol{v}\times\boldsymbol{B})=0,\\
&\nabla \cdot\boldsymbol{B}=0,\\
&\frac{\partial E}{\partial t} + \nabla \cdot \left[\boldsymbol{v}\left(E + P + \frac{B^2}{8\pi}\right) - \frac{\boldsymbol{B}(\boldsymbol{B}\cdot\boldsymbol{v})}{4\pi}\right] = \Gamma - \Lambda +\boldsymbol{f}\cdot \boldsymbol{v} ,
\end{aligned}
\end{equation} 
where $P$ is the thermal pressure, $\boldsymbol{f}$ is a stochastic forcing term applied to drive turbulence, and $E$ is the total energy density. The cooling rate, $\Lambda$, accounts for atomic line cooling and is parameterized as \citep{2002ApJ...564L..97K}:
\begin{equation}
\begin{aligned}
\Lambda = \left(\frac{\rho}{m_{\rm H}}\right)^2 & \left[2\times10^{-19}\exp\left(\frac{-1.148\times 10^5}{T+1000}\right) \right. \\
& \left. + 2.8\times10^{-28}\sqrt{T}\exp\left(\frac{-92}{T}\right)\right] ~~ \mathrm{erg~s^{-1}~cm^{-3}},
\end{aligned}
\end{equation}
where $m_{\rm H}$ is the hydrogen mass. The heating energy density rate, $\Gamma$, is given by: 
\begin{equation}
\Gamma = \left(\frac{2\rho}{m_{\rm H}}\right)\times10^{-26} ~~ \mathrm{erg~s^{-1}~cm^{-3}}.
\end{equation}
We neglect thermal conduction, as its characteristic length scale ($\sim 10^{-2}$~pc) is subdominant on the large scales characteristic of the ISM \citep{2000ApJ...540..271V,2024arXiv240714199H}.

The initial conditions consist of a uniform number density field $n = 1~\mathrm{cm^{-3}}$, corresponding to a mean column density $N_{\rm H} = 3\times10^{20}~\mathrm{cm^2}$, and a uniform magnetic field aligned along the $z$-axis. We adopt a field strength of $B \approx 3$~\SI{}{\micro G}, consistent with Zeeman observations \citep{2012ARA&A..50...29C}. The velocity dispersion is initialized at $\sigma_v \approx 10$~km~s$^{-1}$, based on Larson's law and observational constraints derived from young stellar environments \citep{1981MNRAS.194..809L,2022ApJ...934....7H}. This setup gives a global super-Alfv\'enic regime (e.g., the Alfv\'en Mach number $M_A > 1$), meaning that turbulent magnetic fluctuations dominate the mean magnetic field. Consequently, the statistical properties of the observed structure geometry become largely isotropic and are highly insensitive to the observer's viewing angle.

The simulation box spans 100~pc. To represent the high-latitude ISM, which generally lacks localized, highly compressive drivers like supernova explosions, we employ solenoidal forcing at an injection scale of $L_{\rm inj}=50$~pc, corresponding to a peak wavenumber of $k = 1/L_{\rm inj} = 0.02~\mathrm{pc^{-1}}$. The computational domain is discretized on a uniform $2048^3$ grid, and the effective numerical dissipation scale is approximately 10 cells. We extract four analysis snapshots every 5~Myr after 100 Myr, where 5~Myr is the largest-eddy turnover time. 100 Myr represents the 20 largest-eddy turnover times, ensuring that the turbulence has reached a statistically steady state and the snapshots are independent (see \citealt{2024arXiv240714199H}).

\subsection{Synthetic dust polarization maps}
We use the updated \textsc{Polaris} code from \cite{Giang2023}, incorporating modern physics of grain alignment by radiative torques (RATs) to post-process our MHD simulation. For dust properties, we consider interstellar dust following the Astrodust model (i.e., silicate and carbonaceous components are mixed in one single grain population, see \citealt{Draine.2021}). We use the oblate spheroidal shape of axial ratio $s=1.4$, and the optical constant from the Astrodust model, which was previously constrained from ISM polarimetric observations by \cite{Hensley2023}. The grain size distribution is assumed to follow the MRN power law \citep{Mathis.1977} of $dn/da = C a^{-3.5}$, where $C$ is the normalization constant derived from the dust-to-gas mass ratio of 0.01 and the grain sizes span from the minimum size of $a_{\rm min} = 0.005\,\rm\SI{}{\micro m}$ to the maximum size $a_{\rm max}=0.25~\SI{}{\micro m}$ in the diffuse ISM \citep{Mathis.1977}. The dust composition and size distribution are considered to be uniform across multi-phase ISM components. 

For the dust heating and cooling process, we assume that dust grains are irradiated by the standard diffuse interstellar radiation field (ISRF) from \cite{Mathis.1983} and reemit infrared emission that depends on the dust temperature and optical properties. The dust temperature is calculated in \textsc{Polaris} using Monte Carlo radiative transfer (\citealt{Reissl2016}). The dust grain equilibrium temperature is $T_d \sim 18 - 20$ K in WNM/UNM with $n < 10\,\rm cm^{-3}$, and it decreases slightly to $T_d \sim 16\,\rm K$ in CNM with $n \sim 10 - 100\,\rm cm^{-3}$ because of ISRF attenuation. 

For grain alignment physics, we consider the ideal RAT model in which all grains larger than a critical alignment size, $a_{\rm align}$, determined by local gas conditions and radiation, are perfectly aligned by RATs \citep{2008MNRAS.388..117H}. This ideal RAT model could be achieved for superparamagnetic grains containing embedded iron clusters in the ISM \citep{2007MNRAS.378..910L,HoangLaz.2016,2025ApJ...994..115H}. The calculation of grain alignment properties, including $a_{\rm align}$ and the average alignment efficiency over the grain size distribution $\langle f_{\rm align}\rangle$, is performed with the latest version of the \textsc{Polaris} code (\citealt{Giang2023}). For the multi-phase ISM with $n < 100\,\rm cm^{-3}$, the magnetic alignment by RATs is efficient with small $a_{\rm align} \sim 0.01 - 0.05~\SI{}{\micro m}$ and high average alignment efficiency $\langle f_{\rm align} \rangle \sim 0.75 - 0.85$ (\citealt{HoangTruong.2024}). The fluctuation in the dust polarization is then mainly attributed to the density and magnetic fluctuation effects across ISM-phase components. 

Given dust physical (size distribution, shape, and composition) and grain alignment properties, \textsc{Polaris} solves the polarized radiative transfer in to generate the synthetic thermal dust polarization map at 353 GHz and 150 GHz. For this paper, we analyze the synthetic polarization angles and quantify E/B modes. In the follow-up paper, we will study in detail the effects of multi-phase ISM and magnetic turbulence on the synthetic polarization fraction. 
For more details of the implementations of grain alignment physics and polarized radiative transfer in the \textsc{Polaris} code, please see, e.g., \cite{Reissl2016,Giang2023,Truong.2025}.

\subsection{Structure function} 
We utilize the second-order structure function, a widely accepted method for quantifying the statistical properties of turbulent flows. The structure function of velocity is defined as:
\begin{equation}
\label{eq.sf}
D_v(r)=\langle|\pmb{v}(\pmb{x}+\pmb{r})-\pmb{v}(\pmb{x})|^2\rangle_r,
\end{equation}
where $\pmb{v}(\pmb{x})$ is the velocity at position $\pmb{x}$, and $\pmb{r}$ represents the separation vector. This function can be adapted to measure magnetic field or gas density fluctuations by replacing $\pmb{v}(\pmb{x})$ with $\pmb{B}(\pmb{x})$ or $n(\pmb{x})$, respectively. In the following, we will add the subscripts ``$v$'', ``$B$'', or ``$n$'' to $D(r)$ for distinguishing velocity, magnetic field, and number density accordingly.

To investigate the anisotropy in turbulence within our simulations, we decompose the structure-function into components parallel and perpendicular to the local magnetic fields, following the methodology described in \cite{CV20}:
\begin{equation}
\label{eq.sf_loc}
\begin{aligned}
&\pmb{B}'=\frac{1}{2}(\boldsymbol{B}(\pmb{x}+\pmb{r})+\pmb{B}(\pmb{x})),\\
&D_v(R,z)=\langle|\pmb{v}(\pmb{x}+\pmb{r})-\pmb{v}(\pmb{x})|^2\rangle_r,
\end{aligned}
\end{equation}
where $\pmb{B}'$ defines the local magnetic field direction in a cylindrical coordinate system, with $z$-axis parallel to $\pmb{B}'$, $R=|\hat{z}\times\pmb{r}|$ and $z=\hat{z}\cdot\pmb{r}$, where $\hat{z}=\pmb{B}'/|\pmb{B}'|$.

From these, we can derive the parallel ($\delta v_\parallel$) or perpendicular ($\delta v_\bot$) velocity fluctuations relative to the local magnetic fields as follows:
\begin{equation}
\label{eq.sf_v}
\begin{aligned}
&\delta v_\bot=\sqrt{D_v^\bot}=\sqrt{D_v(R,0)},\\
&\delta v_\parallel=\sqrt{D_v^\parallel}=\sqrt{D_v(0,z)}.
\end{aligned}
\end{equation}

Similarly, the parallel ($\delta B_\parallel$) or perpendicular ($\delta B_\bot$) magnetic field fluctuations are obtained from:
\begin{equation}
\label{eq.sf_b}
\begin{aligned}
\delta B_{\bot}&=\sqrt{D_B^\bot}=\sqrt{D_B(R,0)},\\
\delta B_{\parallel}&=\sqrt{D_B^\parallel}=\sqrt{D_B(0,z)}.
\end{aligned}
\end{equation}

\section{Results}
\label{sec:results}
\subsection{Turbulent and multiphase nature of the ISM}
Traditionally, the ISM is modeled as a multiphase medium in approximate pressure equilibrium, comprising a warm, diffuse component with temperatures $T > 5000$~K (i.e., WNM); an intermediate-density, thermally unstable component with $200 < T < 5000$~K (i.e., UNM); and a cold, dense component with $T < 200$~K (i.e., CNM). The phase definition is physically motivated: along the thermal-equilibrium curve in the \(P\)-\(n\) phase diagram, the thermally stable WNM and CNM branches have \(dP_{\rm eq}/dn>0\), whereas the thermally unstable branch has \(dP_{\rm eq}/dn<0\). The two turning points separating these branches occur at approximately \(T\simeq 5000~{\rm K}\) and \(T\simeq 200~{\rm K}\), which motivates our adopted temperature boundaries for WNM, UNM, and CNM \citep{1977ApJ...218..148M,2003ApJ...587..278W,2011piim.book.....D,2024arXiv240714199H,2025ApJ...986...62H}. This multiphase structure is illustrated in Fig.~\ref{fig:nTv_map}, which displays 2D cross-sections of the number density, temperature, and velocity fields. 

\begin{figure*}[htbp!]
\centering
\includegraphics[width=0.95\textwidth]{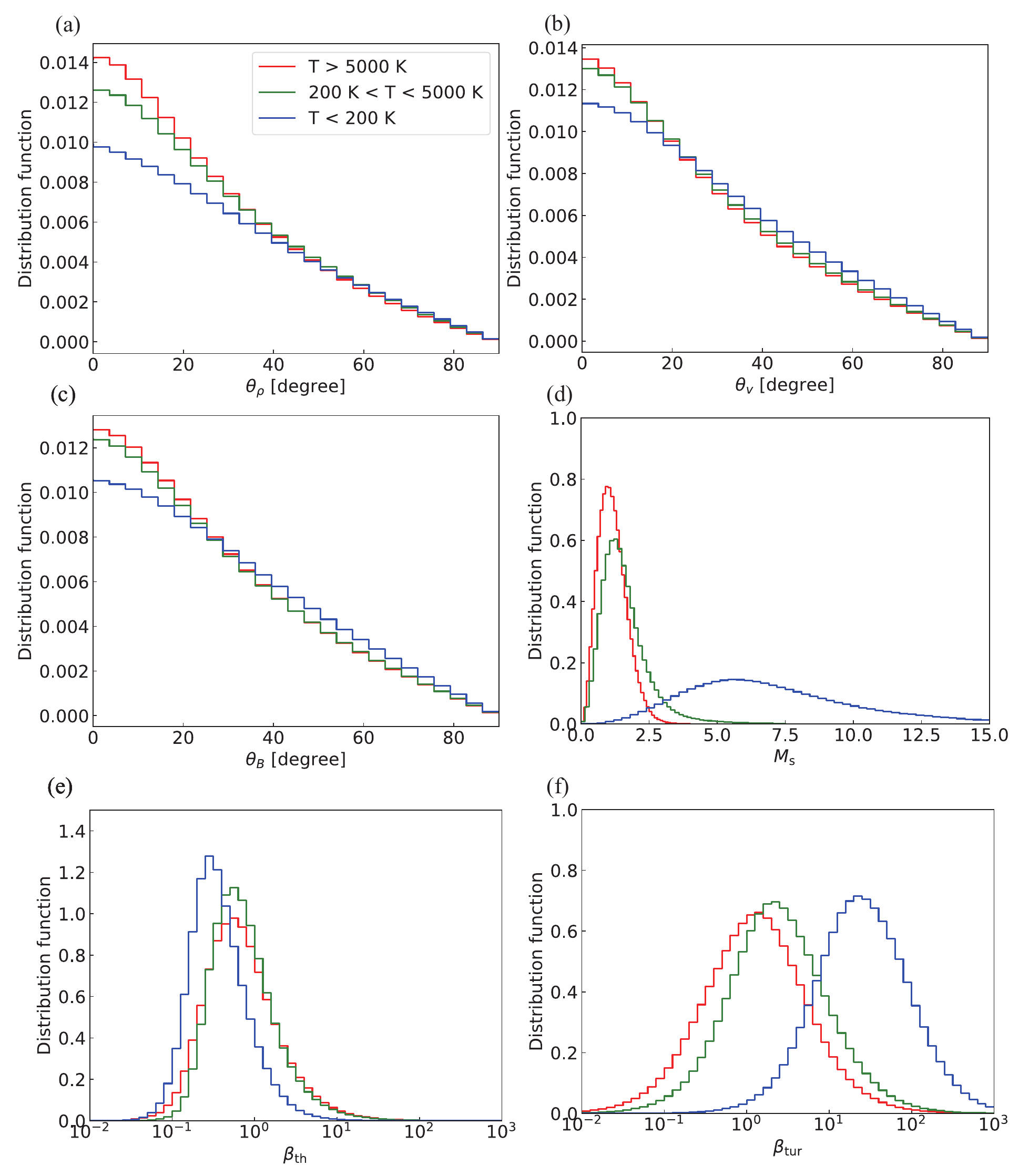}
\caption{\textbf{Panels (a), (b), and (c):} Distributions of the relative angle $\theta_\rho$, $\theta_v$, and $\theta_B$ between the local magnetic field and the rotated (by 90 degrees) density gradient $\nabla\rho$, velocity gradient $\nabla v$, and magnetic field strength gradient $\nabla B$, respectively. The relative angles are evaluated separately for the warm neutral medium (WNM; $T > 5000$ K), unstable neutral medium (UNM; $200$ K $< T <$ 5000 K), and cold neutral medium (CNM; $T < 200$ K). Here, a relative angle of $0^\circ$ corresponds to parallel alignment, while $90^\circ$ indicates perpendicularity. \textbf{Panel (d):} Distribution of the sonic Mach number $M_{\rm s} = v / c_s$ in each phase. \textbf{Panel (e):} Distribution of the thermal plasma beta, $\beta_{\rm th} = (\rho c_s^2/\gamma) / (B^2 / 8\pi)$. \textbf{Panel (f):} Distribution of the turbulent plasma beta, $\beta_{\rm tur} = \rho v^2 / (B^2 / 8\pi)$.  All quantities are phase-separated based on temperature and computed in three dimensions.}
\label{fig:hist_3D}
\end{figure*}

\begin{figure*}[htbp!]
\centering
\includegraphics[width=0.99\textwidth]{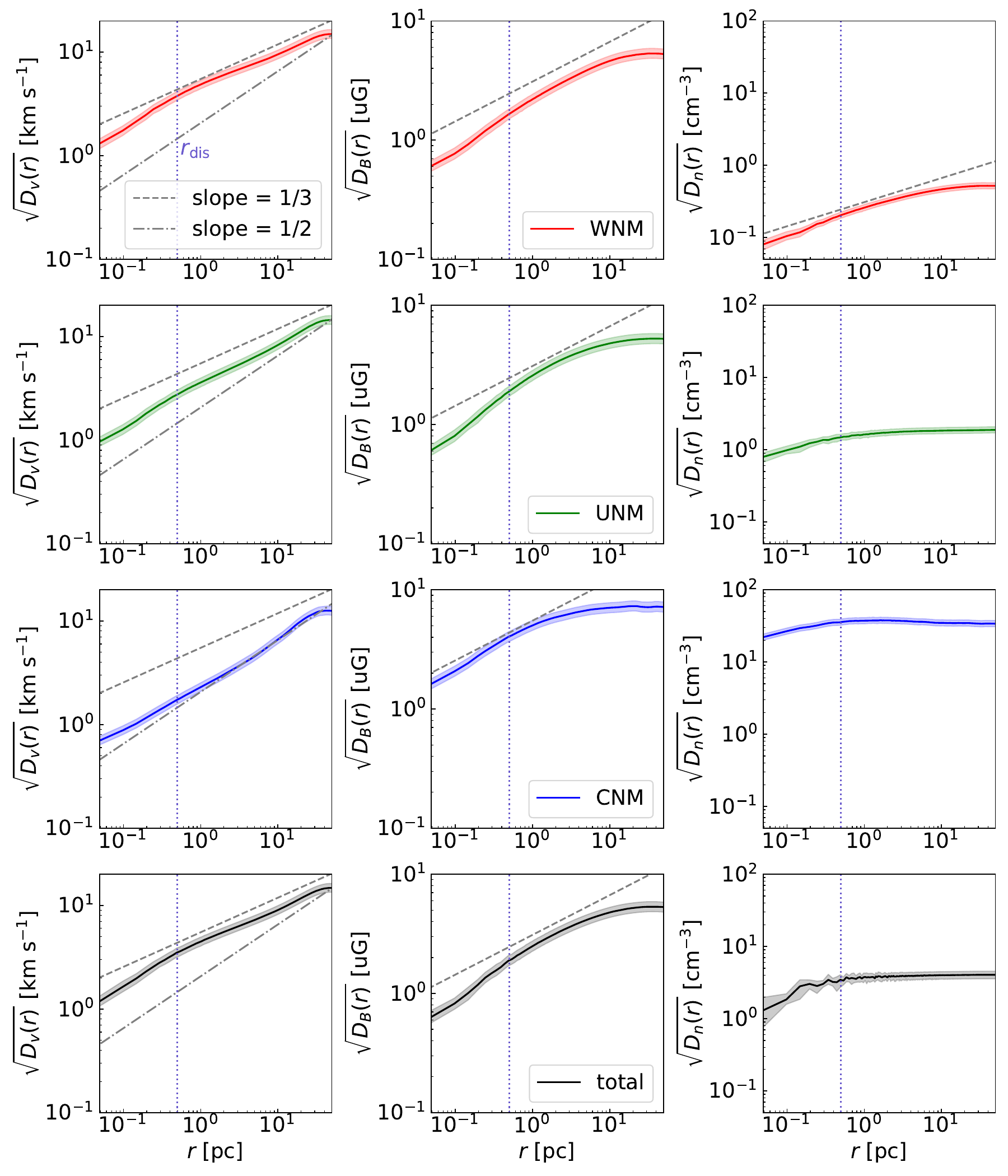}
\caption{Statistical properties of scale-dependent fluctuations of velocity (1st column), magnetic field (2nd column), and number density (3rd column) in different ISM phases. The fluctuations are computed as the square root of the second-order structure function for each quantity, shown separately for the WNM ($T > 5000$ K), UNM ($200$ K $< T <$ 5000 K), CNM ($T < 200$ K), and the total ISM (including all phases). Dashed and dash-dotted reference lines indicate slopes of 1/3 and 1/2, corresponding to Kolmogorov and Burgers turbulence scaling, respectively. The shaded areas represent the standard deviation over the realizations. The numerical dissipation scale $r_{\rm dis}$ is around 0.5 pc. }
\label{fig:sf_wuc}
\end{figure*}

\begin{figure*}
  \centering
  \includegraphics[width=0.82\textwidth]{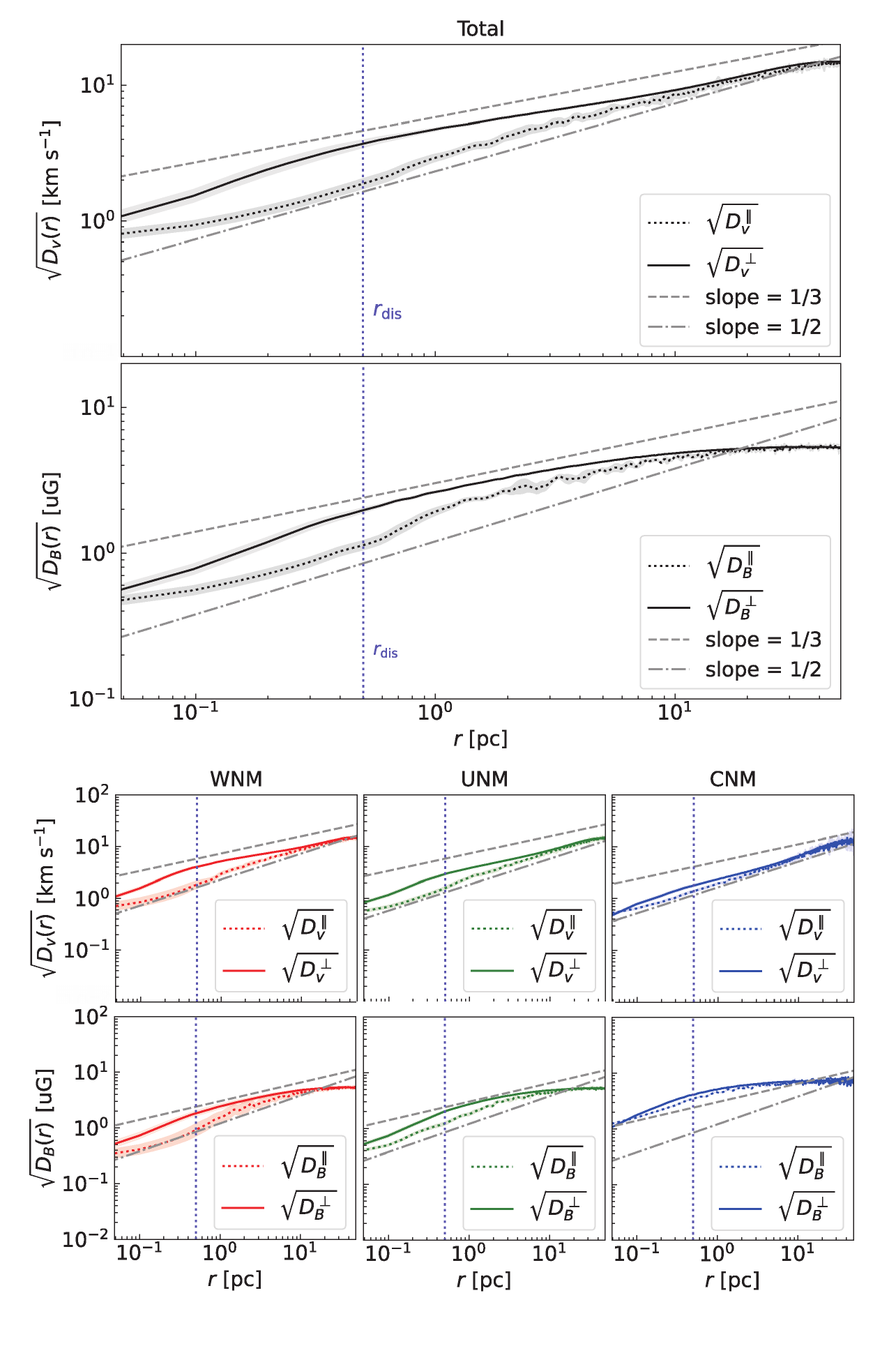}
  \caption{Statistical properties of scale-dependent velocity and magnetic field fluctuations decomposed into components parallel and perpendicular to the local magnetic field in different ISM phases. Fluctuations are computed as the square root of the second-order structure function. They are shown separately for the WNM, UNM, CNM, and the total ISM (including all phases). The shaded areas represent the standard deviation over the realizations. Dashed and dash-dotted grey lines represent reference slopes of 1/3 and 1/2, corresponding to Kolmogorov and Burgers turbulence scalings, respectively. The numerical dissipation scale $r_{\rm dis}$ is around 0.5 pc.}
    \label{fig:SF_decompose}
\end{figure*}
\subsubsection{Thermal, kinetic, and magnetic field energy budget}
Turbulence induces significant fluctuations in density, temperature, and velocity throughout the ISM. In particular, turbulent mixing plays a crucial role in redistributing energy and sustaining the UNM, while thermal instability acts as a secondary contributor \citep{2000ApJ...540..271V,2024arXiv240714199H,2025ApJ...986...62H}. The interplay between turbulent mixing and magnetic fields—both of which support the intermediate phase—results in mass fractions of 21.79\% for the CNM, 33.93\% for the WNM, and 44.28\% for the UNM. The mass fractions are consistent with the multi-phase ISM in high latitude $|b|>30^\circ$ regions \citep{2018A&A...619A..58K}, which are the major targets for the foreground polarization subtraction \citep{2016A&A...586A.133P,2020A&A...641A..11P}.   

The relative importance of thermal pressure and magnetic forces is characterized by the thermal plasma beta, $\beta_{\rm th} = (\rho c_s^2/\gamma) / (B^2 / 8\pi)$, where $c_s$ is the sound speed and $\gamma$ is the adiabatic index. As shown in Fig.~\ref{fig:hist_3D} and summarized in Tab.~\ref{Tab.2}, $\beta_{\rm th}$ in the WNM and UNM spans a broad range from $10^{-1}$ to $10^2$, with a median value $\sim 0.7$. In contrast, the CNM exhibits a narrower distribution ($10^{-1}$ to $10^1$) with a lower median of $0.4$. Dynamically, the WNM and UNM are transonic, with median sonic Mach numbers ($M_s$) of $1.1$--$1.5$. The CNM, however, is highly supersonic with a median $M_s \approx 6.6$, indicating that shock compression plays an important role in shaping this cold phase.

To quantify the balance between turbulent kinetic energy and magnetic energy, we examine the turbulent plasma beta, $\beta_{\rm tur} = \rho v^2 / (B^2 / 8\pi) = 2M_A^2$, where $M_A$ is the Alfv\'en Mach number. In the WNM and UNM, $\beta_{\rm tur}$ ranges from $10^{-2}$ to $10^2$ with median values between $1.4$ and $2.8$. This indicates that these phases are generally in a trans-Alfv\'enic state where magnetic and turbulent kinetic energies are comparable, although local conditions may vary. Conversely, the CNM is characterized by super-Alfv\'enic (i.e., $\beta_{\rm tur}>2$ or $M_A>1$) turbulence; with a median $\beta_{\rm tur} \approx 40$, the turbulent kinetic energy significantly dominates, rendering the magnetic field dynamically subdominant.

The super-Alfv\'enic nature of the CNM is consistent with turbulent dynamo theory \citep{2022MNRAS.514..957S}. Assuming the largest scale of a CNM cloud is $\sim 10$ pc, the dynamo saturation timescale is estimated to exceed ten large-scale eddy turnover times, i.e., $t_{\rm dyn}\sim10t_{\rm eddy}\sim20$ Myr. The supersonic CNM is also affected by shock compression that operates on a shorter timescale of $t_{\rm comp} \sim 2$ Myr. However, turbulent mixing occurs rapidly, with its rate determined by the smallest scales ($t_{\rm mix} \sim 0.3$ Myr at the $\sim 0.5$ pc dissipation scale). It acts to disrupt the compression process. Consequently, these dynamic timescales are significantly shorter than the dynamo saturation timescale, so that the turbulent kinetic energy consistently exceeds the magnetic energy. This dominance of turbulence suggests that the magnetic field within the CNM is highly stochastic and passively tangled by the turbulent motions.

\subsubsection{Parallel alignment of density, velocity, and magnetic field structures with the local magnetic field}
\label{sec:3.1.2}
The variation in the energy partition between turbulence and magnetic fields profoundly influences the geometric alignment of ISM structures. In transonic (or subsonic) compressible MHD turbulence, the turbulent cascade proceeds preferentially in the perpendicular direction, promoting the mixing of density fluctuations into elongated filaments aligned parallel to the local magnetic field \citep{2019ApJ...878..157X, 2025ApJ...986...62H}. The aspect ratio of these filaments is governed by the degree of turbulence anisotropy \citep{2019ApJ...878..157X} and modulated by phase transition processes \citep{2023MNRAS.521..230H}. Conversely, in the super-Alfv\'enic and supersonic regimes, the magnetic field is dynamically subdominant. Here, isotropic shock compression becomes the primary shaping mechanism, capable of generating density structures oriented perpendicular to the magnetic field, e.g., in post-shock layers \citep{2019ApJ...878..157X,2019ApJ...886...17H,2020MNRAS.492..668B}.

In Fig.~\ref{fig:hist_3D}, we quantify this alignment using the relative angles $\theta_\rho$, $\theta_v$, and $\theta_B$. These are defined as the angles between the local magnetic field and the vectors perpendicular to the gradients $\nabla\rho$, $\nabla v$, and $\nabla B$ (i.e., the gradients rotated by $90^\circ$). The gradients are calculated at the numerical dissipation scale ($\sim0.5$~pc). This rotation ensures that the angles represent the orientation of the structure's elongation relative to the local magnetic field: $\theta \approx 0^\circ$ indicates parallel alignment (structures aligned with $B$), while $\theta \approx 90^\circ$ indicates perpendicularity. 

The distributions of $\theta_\rho$ reveal distinct phase-dependent behaviors (see Fig.~\ref{fig:hist_3D}), while all distributions peak at zero. $\theta_\rho$ in WNM has the least dispersed distribution, indicating the most pronounced parallel alignment. This is typically characterized by trans-Alfv\'enic and transonic conditions favorable for anisotropic MHD cascading.
The UNM shows intermediate alignment, while the super-Alfv\'enic and supersonic CNM exhibits significantly weaker parallel alignment, i.e., broader distribution. This trend confirms theoretical expectations: as the Mach number increases and the Alfv\'enic constraint weakens, supersonic shocks disrupt the field-aligned anisotropy, introducing a population of shock-compressed structures perpendicular to the field \citep{2019ApJ...878..157X,2019ApJ...886...17H,2020MNRAS.492..668B}. This phase-dependent density structural alignment has direct consequences for the polarization statistics discussed in \S~\ref{subsec:EB}. The strong parallel alignment in the trans-Alfv\'enic WNM and UNM effectively projects power into the $E$-mode, enhancing the $E$-mode and $B$-mode's power spectral ratio. In contrast, the randomization of structure orientation (or tendency toward perpendicularity) in the shock-dominated CNM reduces the geometric anisotropy, expected to drive the $EE/BB$ ratio closer to unity.

The distributions of $\theta_v$ and $\theta_B$ are similar to that of $\theta_\rho$, showing strong parallel alignment with the local magnetic field. This demonstrates the physical foundations of using velocity gradient \citep{Vel_grad18,2018MNRAS.480.1333H} and synchrotron intensity gradient \citep{synch_grad17,Hu_clusters24} to trace the magnetic field in diffuse ISM.

\subsection{Statistics of velocity, magnetic field, and density fluctuations}
Fig.~\ref{fig:sf_wuc} presents the square root of the undecomposed second-order structure functions for velocity, magnetic field, and density. The quantitative power-law slopes $\alpha$ fitted in the inertial range ($0.5-10$~pc) are summarized in Tab.~\ref{Tab.1}. Rather than performing a single globally averaged fit, we computed the structure functions for four independent realizations and conducted the power-law regression on each realization separately. Consequently, the slopes reported in Tabs.~\ref{Tab.1} and \ref{Tab.2} represent the statistical means of these independent fits, while the quoted uncertainties correspond to the standard deviation across the ensemble. In the subsequent plots, these statistical variations are visualized as shaded regions enveloping the ensemble-mean structure functions.

For velocity, the WNM exhibits a power-law slope of $\alpha_v \approx 0.29$, which is slightly shallower than the standard Kolmogorov turbulence scaling ($1/3$). This slope progressively steepens in the denser phases. In the UNM, the slope increases to $\alpha_v \approx 0.36$. In the CNM, the slope reaches $\alpha_v \approx 0.46$, closely approaching the value of $1/2$ indicative of Burgers turbulence (shock-dominated). While this steepening in the CNM is typically attributed to the presence of supersonic shocks ($M_s \approx 6.6$) that efficiently dissipate kinetic energy, \citet{2025ApJ...986...62H} found that similar steepening can occur in the subsonic WNM under hydrodynamical conditions. This suggests that phase transition dynamics may also play a role in modifying the energy cascade. Finally, the velocity structure function of the total ISM ($\alpha_v \approx 0.31$) falls between that of the volume-filling WNM and UNM ($\alpha_v \approx 0.29$ and $\alpha_v \approx 0.36$), indicating that the WNM and UNM dominate the global turbulent velocity statistics.

For the magnetic field structure function, the WNM exhibits a slope of $\alpha_B \approx 0.33$, perfectly consistent with Kolmogorov scaling. The UNM shows a slightly shallower slope of $\alpha_B \approx 0.28$. In contrast, the CNM displays a significantly shallower magnetic field slope of $\alpha_B \approx 0.15$. This low value, combined with the flattening observed at scales above $5$~pc, suggests that the magnetic field becomes uncorrelated at larger scales within the clumpy cold medium. Despite the high amplitude of fluctuations in the CNM, the structure function of the total ISM ($\alpha_B \approx 0.29$) closely tracks the WNM and UNM values. This implies that, while the CNM contains intense local fluctuations, the globally volume-filling WNM and UNM govern the overall magnetic field statistics of the ISM.

Regarding the density structure function, the UNM and CNM lack a clear power-law scaling. Their slopes are close to zero ($\alpha_n \approx 0.05$ for UNM and $\alpha_n \approx -0.03$ for CNM), indicating that the correlation of density fluctuations in these phases is insignificant and power is concentrated at small scales. The WNM, despite having a lower mean density, exhibits a discernible—albeit shallow—power-law trend with $\alpha_n \approx 0.26$. Notably, unlike the velocity and magnetic field statistics, the density structure function of the total ISM is nearly flat ($\alpha_n \approx 0.03$) and is dominated by the contributions from the dense UNM and CNM phases.

\begin{table*}[htbp!]
    \centering
    \renewcommand{\arraystretch}{1.5}
    \setlength{\tabcolsep}{14pt}
    \caption{Properties of the square root of the second-order structure function in the Multi-phase Medium}
    \begin{tabular}{l c c c c c c c c c}
        \toprule
        \toprule
        \multirow{2}{*}{Phase} & 
        \multirow{2}{*}{$f_{\rm mass}$ (\%)} & 
        \multirow{2}{*}{$f_{\rm vol}$ (\%)} & 
        \multirow{2}{*}{Slope $\alpha_n$} & 
        \multicolumn{3}{c}{Slope $\alpha_v$} & 
        \multicolumn{3}{c}{Slope $\alpha_B$} \\ 
         \cmidrule(lr){5-7} \cmidrule(lr){8-10}
        & & & Tot. & Tot. & $\parallel$ & $\perp$ & Tot. & $\parallel$ & $\perp$ \\ 
        \midrule
        Total & 100   & 100   & 0.03 & 0.31 & 0.47 & 0.29 & 0.29 & 0.39 & 0.27 \\
        WNM   & 33.93 & 61.04 & 0.26 & 0.29 & 0.49 & 0.27 & 0.33 & 0.45 & 0.30 \\
        UNM   & 44.28 & 38.01 & 0.05 & 0.36 & 0.50 & 0.35 & 0.28 & 0.39 & 0.26 \\
        CNM   & 21.79 & 0.95  & -0.03 & 0.46 & 0.52 & 0.45 & 0.15 & 0.19 & 0.15 \\
        \bottomrule
    \end{tabular}
    
    \parbox{\linewidth}{\vspace{1ex}
    \textbf{Notes.} Columns list the mass fraction ($f_{\rm mass}$), volume fraction ($f_{\rm vol}$), and the power-law slope ($\alpha$) derived from the square root of the second-order structure functions, fitted over the range $0.5-10$~pc. The standard deviation across the ensemble of all slopes is $\leq0.04$. Subscripts $v$, $B$, and $n$ denote velocity, magnetic field, and density, respectively. For velocity and magnetic fields, scaling slopes are provided for the total (Tot.) field as well as for components parallel ($\parallel$) and perpendicular ($\perp$) to the local magnetic field.
    }
    \label{Tab.1}
\end{table*}

\begin{figure*}[htbp!]
\centering
\includegraphics[width=0.99\textwidth]{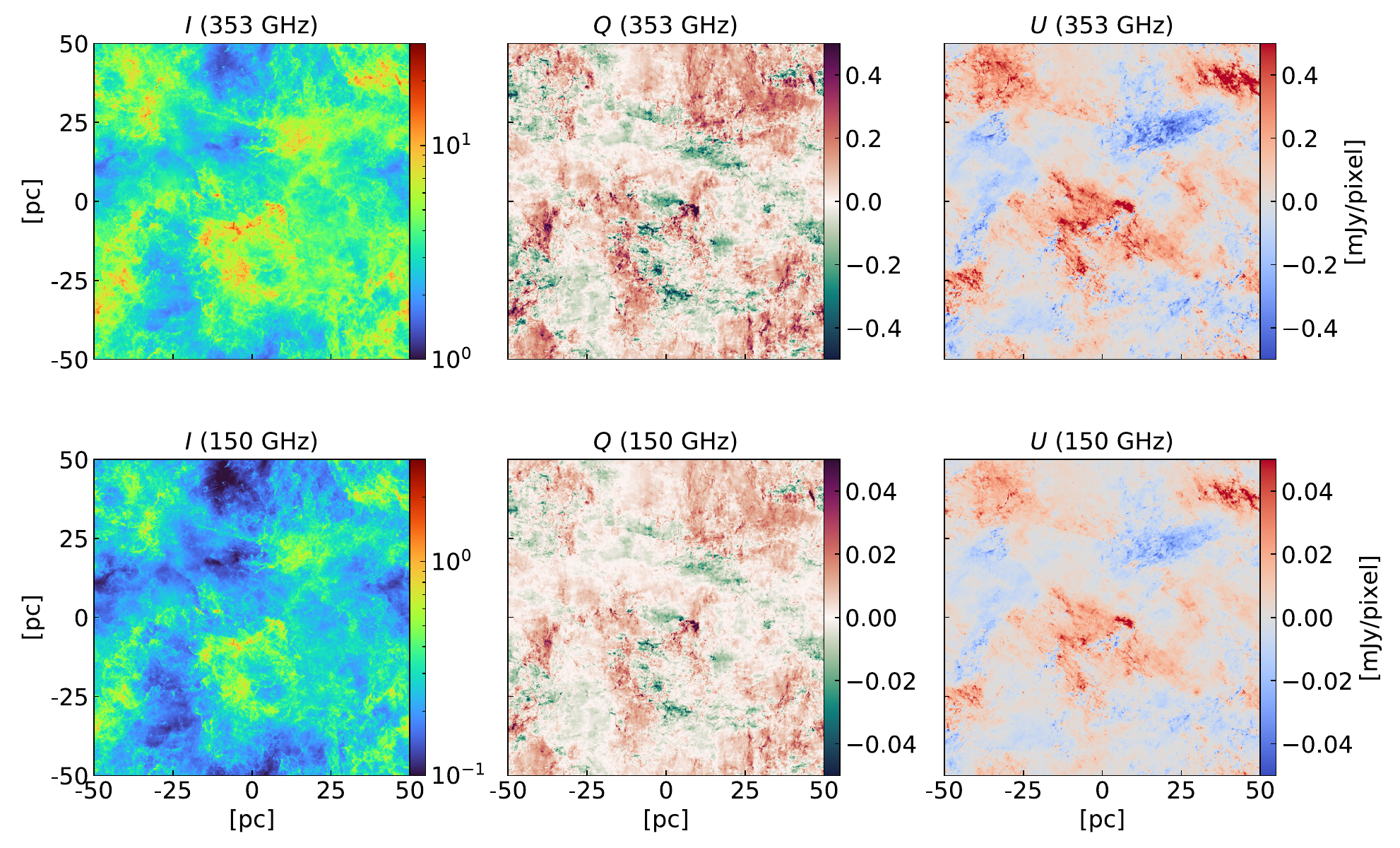}
\caption{Distributions of the Stokes parameters $I$ (left), $Q$ (middle), and $U$ (right) for synthetic dust polarization. We consider two distinct frequencies: 353 GHz (top), corresponding to \textit{Planck} observations, and 150 GHz (bottom), a primary target for future CMB experiments. }
\label{fig:dust_map}
\end{figure*}

\begin{figure*}[p]
\centering
\includegraphics[width=0.99\textwidth]{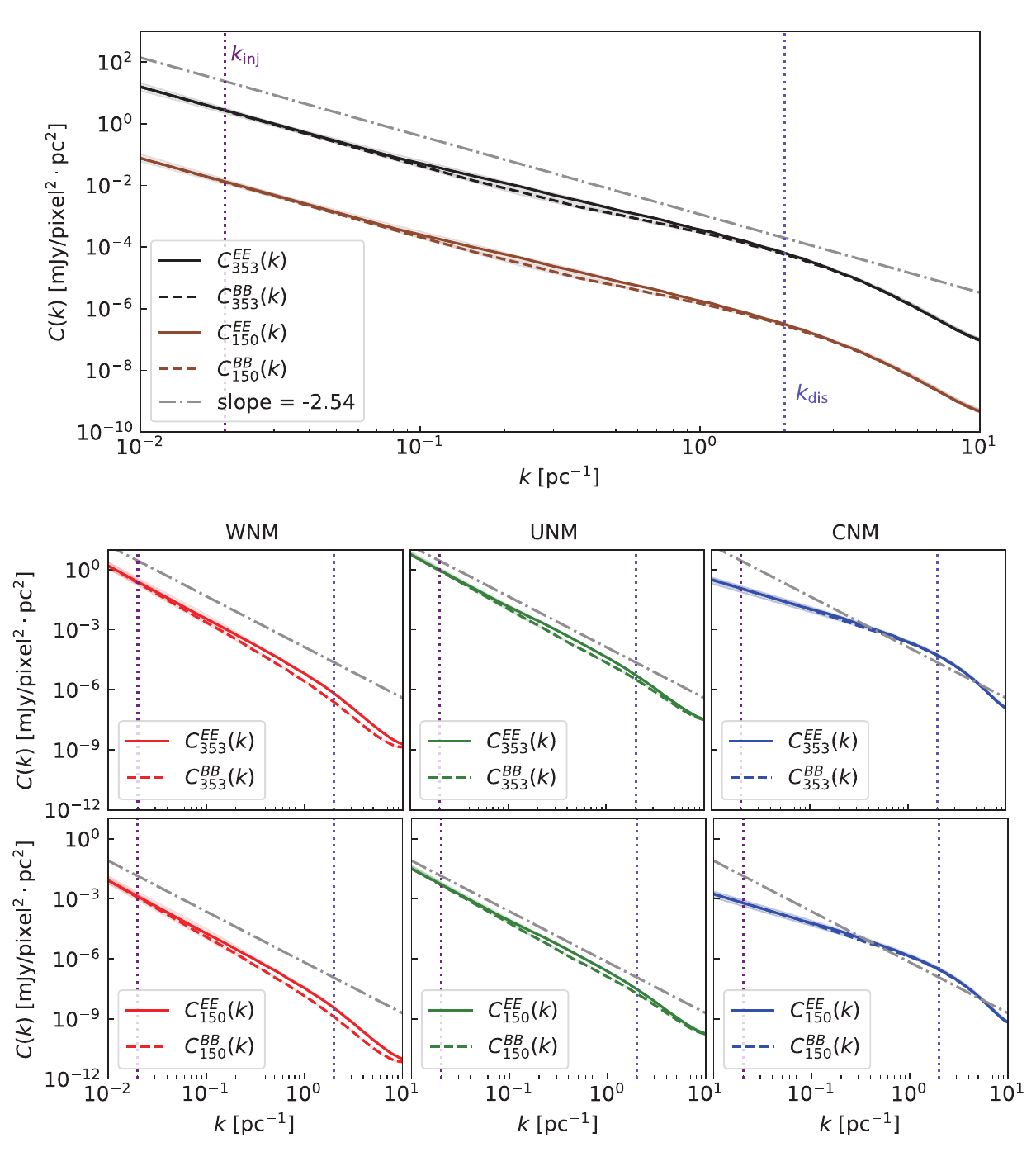}
\caption{$E$-mode and $B$-mode power spectra, $C^{\rm EE}$ and $C^{\rm BB}$, calculated from synthetic dust polarization. \textbf{Top panel:} The power spectra $C^{\rm EE}$ and $C^{\rm BB}$ computed from synthetic dust polarization maps. The gray solid line indicates a reference slope of $-2.54$, which was the $C^{\rm BB}$ spectral slope measured by \textit{Planck} at 353 GHz. The subscripts ``353'' and ``150'' represent two distinct frequencies: 353 GHz and 150 GHz. $k_{\rm inj}$ means the injection wavenumber of turbulence and $k_{\rm dis}$ is the numerical dissipation wavenumber. The shaded area represents the standard deviation. \textbf{Bottom panel:} same as the top panel, but for $C^{\rm EE}$ and $C^{\rm BB}$, calculated from dust polarization in WNM, UNM, and CNM, respectively. }
\label{fig:eb_spectra_x}
\end{figure*}

\begin{table*}[htbp!]
    \centering
    \renewcommand{\arraystretch}{1.5}
    \setlength{\tabcolsep}{6pt}

    \caption{Properties of the Multi-phase Medium and Polarization Statistics}
    
    \begin{tabular}{l c c c c c c c c c c}
        \toprule
        \toprule
        \multirow{2}{*}{Phase} & 
        \multirow{2}{*}{$M_s$} & 
        \multirow{2}{*}{\shortstack{$c_s$ \scriptsize (km s$^{-1}$)}} & 
        \multirow{2}{*}{$\beta_{\rm th}$} & 
        \multirow{2}{*}{$\beta_{\rm tur}$} & 
        \multicolumn{2}{c}{Slope $\alpha^*_{EE}$} & 
        \multicolumn{2}{c}{Slope $\alpha^*_{BB}$} & 
        \multicolumn{2}{c}{Peak $C^{EE}/C^{BB}$} \\
        \cmidrule(lr){6-7} \cmidrule(lr){8-9} \cmidrule(lr){10-11}
         & & & & & 
         353 GHz & 150 GHz & 
         353 GHz & 150 GHz & 
         353 GHz & 150 GHz \\
        \midrule
        Total  & 1.27 & 7.40 & 0.66 & 1.98 & - & - & - & - & 1.52 & 1.52 \\
        WNM   & 1.11 & 8.09 & 0.66 & 1.41 & $-2.99\pm0.01$ & $-2.99\pm0.01$ & $-3.18\pm0.05$ & $-3.18\pm0.05$ & 2.92 & 2.91 \\
        UNM   & 1.45 & 6.38 & 0.70 & 2.82 & $-2.78\pm0.02$ & $-2.78\pm0.02$ & $-2.68\pm0.03$ & $-2.70\pm0.03$ & 2.07 & 2.07 \\
        CNM   & 6.66 & 1.17 & 0.42 & 40.25 & $-1.88\pm0.03$ & $-1.88\pm0.03$ & $-1.81\pm0.05$ & $-1.81\pm0.05$ & 1.29 & 1.29 \\
        \bottomrule
    \end{tabular}
    \parbox{\linewidth}{\vspace{1ex}
    \textbf{Notes.} Columns list the median values for the sonic Mach number ($M_s$), sound speed ($c_s$), thermal plasma beta ($\beta_{\rm th}$), and turbulent plasma beta ($\beta_{\rm tur}$). The spectral slope $\alpha^*$ is fitted within the inertial range ($0.02 \lesssim k \lesssim 2\,\mathrm{pc}^{-1}$), with uncertainties corresponding to the standard deviation across the ensemble. Subscripts $EE$'' and $BB$'' denote the $E$-mode ($C^{EE}$) and $B$-mode ($C^{BB}$) power spectra, respectively.
    }
    \label{Tab.2}
\end{table*}

\begin{figure*}[htbp!]
\centering
\includegraphics[width=0.9\textwidth]{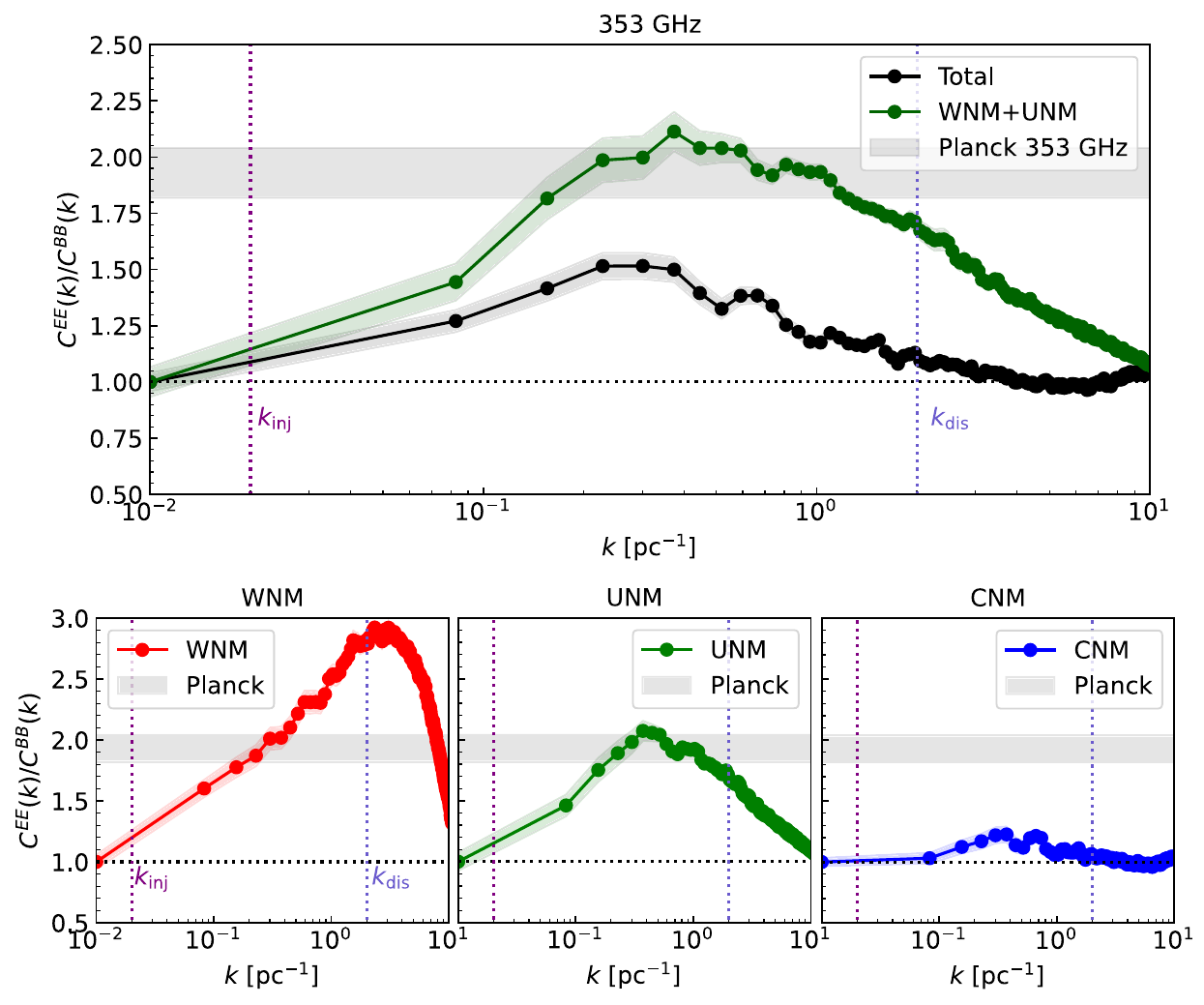}
\caption{Ratio of $C^{\rm EE}$ to $C^{\rm BB}$ from synthetic dust polarization at 353~GHz. \textbf{Top panel:} $EE$ and $BB$ power spectra computed from synthetic dust polarization maps with cold gas ($T < 200$ K) excluded. The shaded area overlapped the line represents the standard deviation. The gray region denotes the range of $C^{\rm EE} / C^{\rm BB}$ ratios measured by \textit{Planck}. $k_{\rm inj}$ means the injection wavenumber of turbulence and $k_{\rm dis}$ is the numerical dissipation wavenumber. \textbf{Bottom panel:} Same as the top panel, but showing spectra computed separately from different gas phases: WNM ($T > 5000$ K, red), UNM ($200$ K $< T <$ 5000 K, green), and CNM ($T < 200$ K, blue).}
\label{fig:eb_ratio}
\end{figure*}
\subsubsection{Anisotropy of the multi-phase ISM}
As shown in Section~\ref{sec:3.1.2}, the gradients of density, velocity, and magnetic field—when rotated by $90^\circ$—are preferentially aligned parallel to the local magnetic field direction. This confirms that the underlying physical density, velocity, and magnetic field structures are themselves elongated parallel to the local magnetic field. Such MHD turbulence anisotropy has profound implications for the alignment, especially between density structures and the magnetic field, In subsonic and transonic compressible MHD turbulence, the turbulent cascade develops preferentially in directions perpendicular to the magnetic field. This anisotropic mixing promotes the formation of elongated density structures that align parallel to the local magnetic field lines \citep{2019ApJ...878..157X}. It consequently can regulate polarization statistics, such as the $EE/BB$ ratio \citep{2018MNRAS.478..530K,2020ApJ...899...31H}. 

In Fig.~\ref{fig:SF_decompose}, we quantify this anisotropy by decomposing the square root of the second-order velocity structure function into components parallel ($\parallel$) and perpendicular ($\perp$) to the local magnetic field. For the total velocity field, the perpendicular component exhibits a slope of $\alpha_{v, \perp} = 0.29$, close to the Kolmogorov scaling, while the parallel component scales with a slope of $\alpha_{v, \parallel} = 0.47$ (see Tab.~\ref{Tab.1}). Crucially, the velocity anisotropy—defined as the ratio of these components—is scale-dependent and becomes more pronounced at smaller scales, consistent with theoretical expectations for magnetized turbulence \citep{LV99,CV20,2021ApJ...911...37H}.

Phase decomposition reveals that the perpendicular component dominates the power across most scales in the diffuse phases, converging with the parallel component only at the driving scale ($\sim 50$~pc). This anisotropy is most prominent in the WNM and UNM. In the WNM, the perpendicular velocity slope is $\alpha_{v, \perp} = 0.27$, while the parallel component reaches $\alpha_{v, \parallel} = 0.49$. The UNM shows a similar trend with $\alpha_{v, \perp} = 0.35$ and $\alpha_{v, \parallel} = 0.50$. In contrast, the super-Alfv\'enic CNM exhibits nearly isotropic scaling, with both perpendicular ($\alpha_{v, \perp} = 0.45$) and parallel ($\alpha_{v, \parallel} = 0.52$) components approaching the $1/2$ slope. Analysis of the magnetic field structure functions (Fig.~\ref{fig:SF_decompose}) shows analogous behavior. The perpendicular magnetic component generally dominates, scaling with a total slope of $\alpha_{B, \perp} = 0.27$, though this value reaches $\alpha_{B, \perp} = 0.30$ in the WNM up to a phase-dependent saturation scale. Beyond this scale—approximately 8~pc in the WNM, 5~pc in the UNM, and 1~pc in the CNM—the slope flattens, indicating reduced correlation in the magnetic fluctuations. The parallel magnetic component in the WNM exhibits a slope of $\alpha_{B, \parallel} = 0.45$, whereas it flattens significantly in the UNM ($0.39$) and CNM ($0.19$). For both velocity and magnetic fields, the perpendicular-to-parallel ratio reaches a factor of $\sim 3$ at the dissipation scale ($0.5$~pc) in the WNM and UNM, while remaining notably lower ($\sim 1.25$) in the CNM.


\begin{figure*}
\centering
\includegraphics[width=0.9\textwidth]{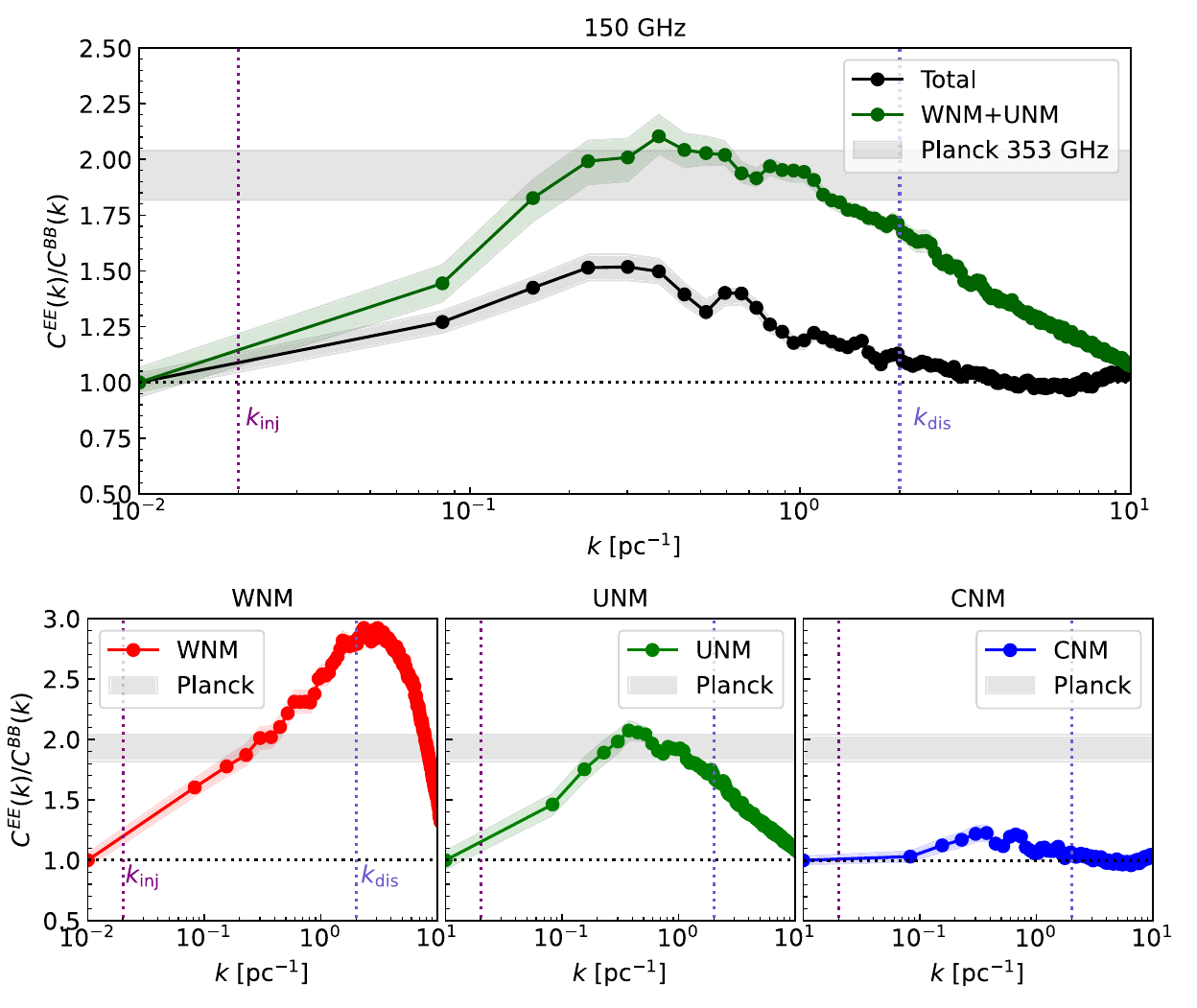}
\caption{Same as Fig.~\ref{fig:eb_ratio}, but at 150~GHz.}
\label{fig:eb_ratio_150}
\end{figure*}

\subsection{Synthetic dust polarization at 150 GHz and 353 GHz} 
\label{subsec:EB}
By post-processing the multi-phase simulation with the radiative transfer code \textsc{Polaris}, we produced synthetic dust polarization maps, including the intensity $I$ and Stokes parameters $Q$ and $U$ at 150 GHz and 353 GHz. The resulting $I$, $Q$, and $U$ maps are presented in Fig.~\ref{fig:dust_map}. At a given frequency, the three maps exhibit distinct spatial morphologies. When comparing the two frequencies, the most notable difference is in amplitude: the 150 GHz maps are approximately one order of magnitude fainter than the 353 GHz maps. This difference arises because thermal dust emission scales rapidly with frequency. Furthermore, while the maps at the two frequencies are globally similar in spatial distributions, significant local variations exist. 

\subsubsection{Power spectra of $E$-mode and $B$-mode}
We compute the power spectra $C^{EE}$ and $C^{BB}$ of $E$-mode and $B$-mode polarization assuming a flat 100~pc patch of the sky, following the flat-sky approximation method \citep[e.g.,][]{2018PhRvL.121b1104K}. We first perform a 2D Fast Fourier Transform (FFT) on the spatial Stokes parameters $Q(\pmb{x})$ and $U(\pmb{x})$ to obtain their Fourier coefficients $\tilde{Q}(\pmb{k})$ and $\tilde{U}(\pmb{k})$. These coefficients are then decomposed into $E$- and $B$-modes via a coordinate rotation in Fourier space:
\begin{equation}
\begin{aligned}
\tilde{E}(\pmb{k}) &= \tilde{Q}(\pmb{k}) \cos 2\phi_{\pmb{k}} + \tilde{U}(\pmb{k}) \sin 2\phi_{\pmb{k}},\\\\
\tilde{B}(\pmb{k}) &= -\tilde{Q}(\pmb{k}) \sin 2\phi_{\pmb{k}} + \tilde{U}(\pmb{k}) \cos 2\phi_{\pmb{k}},
\end{aligned}
\end{equation}
where $\phi_{\pmb{k}} = \arctan(k_y/k_x)$ is the azimuthal angle of the wavevector $\pmb{k}$. Finally, the 1D power spectra $C^{EE}(k)$ and $C^{BB}(k)$ are calculated by azimuthally averaging the squared magnitudes $|\tilde{E}(\pmb{k})|^2$ and $|\tilde{B}(\pmb{k})|^2$ within concentric annular bins, normalized by the physical area of the simulation domain. 

As shown in the top panel of Fig.~\ref{fig:eb_spectra_x}, the $B$-mode power spectrum $C^{BB}$ at 353 GHz at low wavenumbers ($k \le 0.5\,\text{pc}^{-1}$, corresponding to spatial scales $\ge 20\,\text{pc}$) is close to the reference slope of $-2.54$ reported by \textit{Planck} \footnote{It is important to emphasize that the \textit{Planck} high-latitude measurements encompass a vastly different physical volume and geometry compared to our idealized 100~pc domain. Therefore, a direct quantitative match is neither expected nor required to validate our underlying physical conclusions.}, a distinct flattening emerges at $k \gtrsim 0.5\,\text{pc}^{-1}$. Physically, this scale $\le20$~pc likely marks the transition from the diffuse WNM to the clumpy CNM formed. 

To identify the origin of this flattening, we decompose the total polarization into contributions from the WNM, UNM, and CNM phases. This is achieved by masking the density field based on local phase temperatures during the generation of synthetic Stokes parameter maps. Because the far-infrared dust emission is optically thin, this approach physically isolates the true 3D emission contribution of the cold gas without generating 2D boundary artifacts. As illustrated in the middle panel of Fig.~\ref{fig:eb_spectra_x}, the $C^{BB}$ spectra of the individual phases exhibit markedly different characteristics. The WNM spectrum shows a steep power-law slope of $-3.18$, while the UNM spectrum, with a slope of $-2.68$, closestly aligns with the \textit{Planck} reference value. In contrast, the CNM spectrum is significantly flatter, with a slope of $-1.81$. This spectral flatness ensures that the CNM dominates the total power at high wavenumbers, thereby driving the flattening feature observed in the total, undecomposed spectrum. A similar trend is observed for the $E$-mode spectrum $C^{EE}$, which also exhibits small-scale flattening. Notably, $E$-mode power consistently exceeds $B$-mode power ($C^{EE} > C^{BB}$) in the WNM and UNM phases, whereas the two become nearly comparable ($C^{EE} \approx C^{BB}$) in the CNM. It should be noted that although the CNM has a small volume filling fraction, and we used a hard temperature threshold as the phase definition, the high-resolution simulation ensures that the CNM signal is not dominated by shot noise, see the cross-spectra test in Appendix~\ref{app:cross_spectrum}.

These spectral features remain consistent between 150 GHz and 353 GHz. As illustrated in Fig.~\ref{fig:eb_spectra_x}, the power spectra at both frequencies exhibit remarkably similar small-scale flattening and slope transitions, despite the order-of-magnitude difference in their total amplitudes. While our simulation incorporates spatial variations in dust temperature—ranging from approximately $16$ to $20\text{ K}$—the intrinsic dust properties are kept spatially uniform. The frequency-independent nature of the power spectrum shape suggests that these thermal fluctuations do not introduce significant spectral differences. Consequently, the observed flattening is a direct result of the density fluctuations within the CNM rather than the radiative effects. However, power spectra primarily reflect averaged information globally. Locally, radiative transfer effects may still be significant, leading to frequency-dependent differences in dust emission (see Fig.~\ref{fig:dust_map}), particularly in regions where multiphase clouds overlap along the line of sight.

\subsubsection{$EE/BB$ asymmetry in WNM, UNM, and CNM}
Fig.~\ref{fig:eb_ratio} displays the power spectrum ratio $C^{EE}/C^{BB}$ across the sampled scales. A prominent feature within the inertial range ($0.02 \lesssim k \lesssim 2\,\mathrm{pc}^{-1}$) is that the ratio consistently exceeds unity, which is insensitive to beam size smaller than the numerical dissipation scale (see Appendix~\ref{app:beam}). This dominance of $E$-mode power is a characteristic signature of magnetized anisotropic structure in the ISM. Physically, this asymmetry arises from the preferred alignment of dust filaments with local magnetic field lines. Such structural alignment creates a geometric anisotropy that projects significantly more power into $E$-modes than $B$-modes upon LOS integration. Additionally, \cite{2019ApJ...880..106K} found the $EE/BB$ ratio can exhibit temporal fluctuations in the presence of bursty star formation and stellar feedback; our numerical setup is designed as a clean, controlled test of MHD turbulence. Once such turbulence fully develops, its statistical properties remain globally stable. The relevant time scale is 2 - 4 the large-eddy turnover time at the driving scale \citep{2024arXiv240714199H}, which is $\simeq 20~{\rm Myr}$ in our simulations. 

Our total emission results are qualitatively close to \textit{Planck} observations, which reported a ratio of $EE/BB \approx 2$ (indicated by the gray shaded region). Although our peak value \footnote{We focus on the peak ratio due to strict numerical limitations inherent to MHD simulations. Turbulence is driven isotropically at the injection scale (forcing the ratio to 1), and numerical dissipation forces it back to 1 at small scales. Consequently, the physical $EE/BB$ asymmetry can only manifest over a narrow numerical inertial range.}($\sim 1.5$) is lower than the \textit{Planck} value; this difference is primarily attributable to the presence of supersonic and super-Alfv\'enic CNM. Similar reduced ratios ($\sim 1.4–1.7$) have been noted in previous numerical studies \citep{2018PhRvL.121b1104K, 2019ApJ...880..106K,2025PhRvD.112j1302H}. Insufficient resolution was often cited as a potential cause, because at lower resolutions, unresolved cooling lengths in the dense gas lead to artificial numerical fragmentation and sparsity. However, our resolution study in the Appendix~\ref{app:res} demonstrates that the CNM’s contribution to the $EE/BB$ ratio is well-converged at $2048^3$ resolution. Consequently, we conclude that the lower $EE/BB$ ratio and the associated spectral flattening are physical signatures of shocked, supersonic, and super-Alfv\'enic turbulence within the CNM phase.

The multi-phase decomposition further elucidates these individual contributions: the WNM polarization yields a high $EE/BB$ ratio peaking at $2.92$, while the UNM produces a maximum ratio of $\approx 2.07$, broadly consistent with, and closest among the three phases to, the \textit{Planck} -inferred EE/BB ratio. In contrast, the dense CNM yields a significantly lower peak ratio of $\approx 1.29$. Interestingly, excluding the CNM contribution (i.e., considering only the UNM and WNM) results in an aggregate $EE/BB$ ratio of $\approx 2$, a value that persists toward large wavenumbers where the WNM ratio itself exceeds two. Despite its very high ratio at certain wavenumbers, the overall contribution of the WNM to the total power is relatively minor, likely due to its lower density and consequently smaller contribution to the integrated polarization signal. These findings suggest that the UNM, which has the largest mass fraction, may be an important contributor to the polarized dust foreground typically observed at high Galactic latitudes.

Furthermore, the $EE/BB$ trends at 150 GHz and 353 GHz exhibit negligible differences, confirming that this asymmetry is fundamentally driven by the MHD turbulence and remains insensitive to emission frequency in an environment with uniform dust properties. If future observations were to reveal a frequency-dependent $EE/BB$ ratio, it would indicate significant spatial variations in dust intrinsic properties or temperatures. 

\section{Discussion}
\label{sec:discussion}
\subsection{Comparison with earlier work}
While building upon foundational studies of multiphase ISM polarization, our work introduces critical advancements in resolved physical scales and phase-separated turbulence diagnostics. Previous analytical work by \citet{2018MNRAS.478..530K}, assuming isothermal MHD turbulence, predicts that a globally sub-Alfv\'enic regime ($M_A \sim 0.5$) is required to reproduce the \textit{Planck} value of $EE/BB \approx 2$, whereas trans- or super-Alfv\'enic regimes would result in $EE/BB \approx 1$. Our work, however, introduces the turbulence analysis for each phase and demonstrates that global magnetization parameters mask phase-specific behaviors. We show that within a globally super-Alfv\'enic multiphase ISM, the UNM and WNM are actually trans-Alfv\'enic, naturally generating $EE/BB \approx 2$ and $EE/BB > 2$, respectively. Furthermore, the super-Alfv\'enic CNM yields $EE/BB \approx 1.29$, exceeding the symmetric unity prediction. 


Lower-resolution synthetic observations of the multiphase ISM have struggled to match \textit{Planck}, yielding integrated $EE/BB$ ratios that peak only around $\sim 1.4-1.7$ \citep[e.g.,][]{2018PhRvL.121b1104K,2019ApJ...880..106K}. Insufficient resolution was often cited as a potential explanation. By utilizing a $2048^3$ grid, we fully resolve the dense CNM and reach resolution convergence (see Appendix~\ref{app:res}). More importantly, our phase decomposition clearly establishes that the super-Alfv\'enic CNM exhibits poorer 3D magnetic alignment, and its polarization signal suppresses the global $EE/BB$ ratio. Isolating the combined WNM and UNM recovers a ratio ($EE/BB \approx 2.05$) that mimics \textit{Planck} observations. These insights are fundamentally important for interpreting and modeling the Galactic foreground polarization \citep{2017A&A...601A..71G,2020ApJ...899...31H,2022A&A...663A.175K,2025arXiv251211656B}.

\subsection{Implications for frequency decorrelation}

In our \textsc{Polaris} radiative transfer modeling, local dust temperatures and grain alignment properties are self-consistently calculated based on the local 3D physical conditions and radiation field. Due to the diffuse nature of the environment, the attenuation of the ISRF is insignificant, resulting in a narrow range of local dust temperature. Moreover, the weak fluctuation of the local gas density combined with a narrow dust temperature range induces a small fluctuation in the critical alignment size $a_{\rm align}$, which does not change the polarization spectrum considerably at long submm wavelengths. It suggests that the variations in thermodynamic temperature naturally arising from multiphase MHD turbulence and grain alignment efficiencies (e.g., via RAT) are insufficient to drive the frequency decorrelation. This implies that explaining the Galactic foreground decorrelation may require spatial variations in intrinsic dust properties, such as grain size distribution and/or dust composition, which are not available in the Astrodust model \citep{Hensley2023}. A detailed analysis of polarization properties regarding the overlap of multiple independent clouds along a single line of sight is presented in {\color{BlueViolet}Truong et al. (2026)}.

Other forms of alignment torques, such as mechanical alignment torques (METs) arising from gas--grain interactions, are not expected to qualitatively change this conclusion. Under typical diffuse-ISM conditions, METs are expected to align grains with their long axes perpendicular to the magnetic field \citep{2007MNRAS.378..910L,2018ApJ...852..129H}, similar to radiative alignment torques. Thus, the polarization orientation remains primarily determined by the magnetic-field geometry. METs may enhance the overall alignment efficiency, for example, by reducing the minimum grain size that can be efficiently aligned ($a_{\rm align}$). Including METs would therefore likely increase the polarization fraction by allowing a larger fraction of grains to contribute to polarized emission. However, because the polarization angle would still be controlled mainly by the magnetic-field direction, the morphology of the polarization pattern is not expected to change significantly. Since our analysis and conclusions are based primarily on polarization-angle statistics and the resulting $EE/BB$ ratio, we do not expect the inclusion of MET alignment to qualitatively affect our results.

\section{Conclusion}
\label{sec:conclusion}
In this study, we investigated the physical origin of Galactic dust polarization properties using high-resolution ($2048^3$) three-dimensional MHD simulations of the turbulent multiphase ISM, coupled with synthetic dust polarization calculations using the \textsc{Polaris} radiative-transfer code. By separating the warm neutral medium (WNM), unstable neutral medium (UNM), and cold neutral medium (CNM), we quantified how the distinct turbulence regimes, magnetic-field structures, and density--field alignments in each phase shape polarized thermal dust emission at 353 GHz and 150 GHz. Our main findings are summarized as follows:

\begin{enumerate}
\item Under typical diffuse ISM conditions, with $(B\simeq 3~\SI{}{\micro \rm G})$ and $(\sigma_v\simeq 10~{\rm km~s^{-1}})$, the WNM and UNM are globally transonic and trans-Alfv\'enic, with ($M_s\sim 1.1-1.5$) and ($\beta_{\rm tur}\sim 1.4-2.8$). These conditions favor anisotropic velocity and magnetic-field fluctuations, with perpendicular fluctuations relative to the local magnetic field exceeding the parallel components. In contrast, the CNM is highly supersonic and super-Alfv\'enic, with ($M_s\simeq 6.6$) and ($\beta_{\rm tur}\simeq 40$). In this regime, turbulent motions dominate over magnetic stresses, making the CNM magnetic field more stochastic and passively tangled by the flow.

\item The WNM, UNM, and CNM exhibit distinct statistical scalings of turbulent fluctuations. The square root of the second-order velocity structure function follows a Kolmogorov-like slope, close to \(1/3\), in the WNM and UNM, consistent with subsonic/transonic turbulence. In the CNM, the velocity scaling steepens to \(\sim 1/2\), indicative of shock-dominated Burgers-like turbulence. Magnetic-field fluctuations in the WNM approximately track the velocity scaling, whereas the CNM magnetic field shows a much shallower scaling and becomes weakly correlated on scales larger than a few pc. The density structure functions of the dense UNM and CNM are nearly flat, indicating that density fluctuations in these phases are concentrated primarily at small scales.

\item The spectral slopes of the \(E\)- and \(B\)-mode power spectra vary significantly among the different ISM phases. The UNM dust component yields spectral slopes closest to the Planck-inferred values at 353 GHz, whereas the WNM contribution produces steeper spectra. In contrast, the CNM produces significantly shallower spectra, with noticeable small-scale flattening at \(k\gtrsim 0.5~{\rm pc^{-1}}\). This flattening reflects the clumpy, shock-compressed nature of the CNM and its relatively weak magnetic-field alignment.

\item Comparing synthetic polarization maps at 150 GHz and 353 GHz, we find that the power spectra show similar phase-dependent slopes and small-scale flattening at both frequencies. These nearly frequency-independent features indicate that, in our model, the scale-dependent polarization statistics are primarily controlled by the spatial structure of the density and magnetic field, especially the CNM density fluctuations, rather than by temperature variations or frequency-dependent radiative-transfer effects.

\item The trans-Alfv\'enic WNM and UNM exhibit tight parallel alignment between density structures and the local magnetic field. This alignment enhances the E-mode power and yields high \(EE/BB\) polarization ratios, peaking at \(\approx 2\) for the UNM and \(\approx 3\) for the WNM. Conversely, the super-Alfv\'enic and supersonic CNM shows weaker density--field alignment and a more stochastic magnetic-field geometry, resulting in a lower \(EE/BB\) ratio of \(\sim 1.3\), closer to equal E- and B-mode power. When the CNM contribution is excluded, the combined WNM+UNM emission yields \(EE/BB\approx 2\), broadly consistent with the Planck-inferred high-latitude value at 353 GHz. This suggests that diffuse WNM/UNM gas, and particularly the mass-dominant UNM in our simulations, may provide an important contribution to the polarized Galactic dust foreground.
\end{enumerate}



\begin{acknowledgments}
We thank Alex Lazarian for the helpful discussion. Y.H. is supported by the Sherman Fairchild Postdoctoral Fellowship at the California Institute of Technology. Y.H. acknowledges the support for this work provided by NASA through the NASA Hubble Fellowship grant No. HST-HF2-51557.001 awarded by the Space Telescope Science Institute, which is operated by the Association of Universities for Research in Astronomy, Incorporated, under NASA contract NAS5-26555. This work used SDSC Expanse CPU and NCSA Delta CPU and GPU through allocations PHY230032, PHY230033, PHY230091, PHY230105,  PHY230178, and PHY240183, from the Advanced Cyberinfrastructure Coordination Ecosystem: Services \& Support (ACCESS) program, which is supported by National Science Foundation grants \#2138259, \#2138286, \#2138307, \#2137603, and \#2138296.  We acknowledge the computational resources supported by NASA High-End Computing (HEC), Request (SMD-24-33263831). T.H. and B.T. are supported by the main research project (No. 2025186902) from Korea Astronomy and Space Science Institute (KASI). B.T. acknowledges the use of the gnmnu HPC cluster of KASI for performing synthetic modeling of thermal dust polarization and data analysis. T.H. acknowledges the support from the Vietnam National Foundation for Science and Technology Development (NAFOSTED) under grant number 103.99-2024.36. 
\end{acknowledgments}

%

\vspace{10mm}
\software{AthenaK \citep{2024arXiv240916053S}, Python3 \citep{10.5555/1593511}, ChatGPT \citep{gpt}
          }

\newpage
\appendix
\section{Effect of beam size}
\label{app:beam}
\begin{figure*}[htbp!]
\centering
\includegraphics[width=0.85\textwidth]{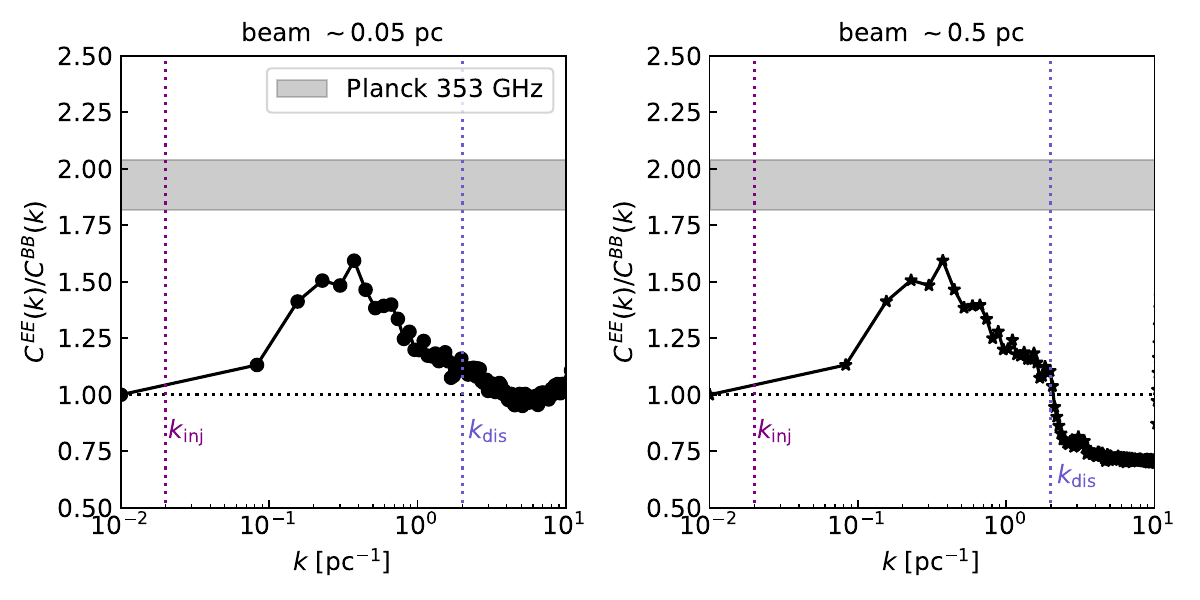}
\caption{Ratio of $C^{\rm EE}$ to $C^{\rm BB}$ from synthetic dust polarization with all phases included. Two different beam sizes $\sim0.05$ pc (left) and $\sim0.5$~pc (right) are included. $k_{\rm inj}$ means the injection wavenumber of turbulence and $k_{\rm dis}$ is the numerical dissipation wavenumber.}
\label{fig:beam}
\end{figure*}

To evaluate the impact of finite angular resolution on our statistics, we compare the power spectra derived using two distinct beam sizes: $\sim 0.05$~pc (close to the grid resolution) and $\sim 0.5$~pc (simulating a coarser observational beam). As illustrated in Fig.~\ref{fig:beam}, the beam smoothing primarily suppresses the power at small scales (high wavenumbers, $k \gtrsim \pi/\theta_{\rm beam}$), acting effectively as a low-pass filter. Crucially, however, the spectral behaviors and the $EE/BB$ ratio at large scales ($k < k_{\rm beam}$) remain unaffected. 

\begin{figure*}[htbp!]
\centering
\includegraphics[width=0.8\textwidth]{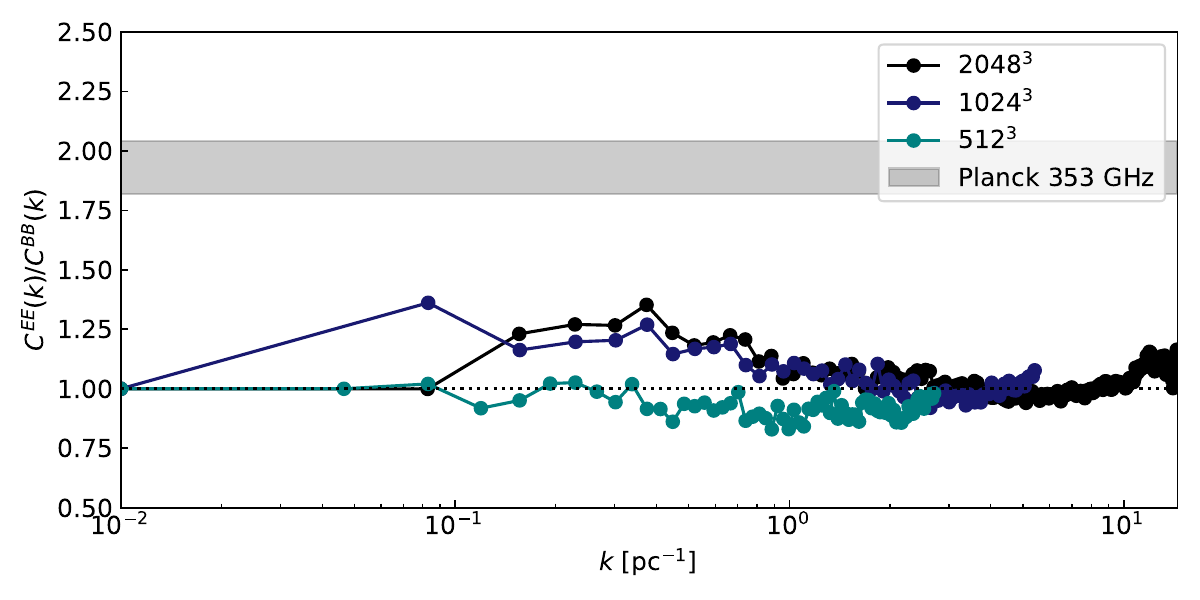}
\caption{Ratio of $C^{\rm EE}$ to $C^{\rm BB}$ for synthetic 353 GHz dust polarization from CNM. Three different numerical grid resolutions $512^3$, $1024^3,$ and $2048^3$ are included.}
\label{fig:resolution}
\end{figure*}

\section{Resolution convergence}
\label{app:res}
To validate the robustness of the low $EE/BB$ ratio observed in the CNM, we perform a resolution study using three numerical grids: $512^3$, $1024^3$, and $2048^3$. Fig.~\ref{fig:beam} displays the ratio of $C^{\rm EE}$ to $C^{\rm BB}$ for synthetic dust polarization from CNM at 353 GHz. While the lowest resolution run ($512^3$) slightly underestimates the ratio, the $1024^3$ and $2048^3$ simulations show convergence, particularly in the inertial range ($0.02\lesssim k \lesssim 2\,\mathrm{pc}^{-1}$). Crucially, even at the highest resolution, the maximum ratio is close to 1.3 and distinct from the \textit{Planck} 353 GHz value of $\sim 2$.


\section{Cross-spectrum test for CNM sparsity artifacts}
\label{app:cross_spectrum}

\begin{figure*}[htbp!]
\centering
\includegraphics[width=0.5\textwidth]{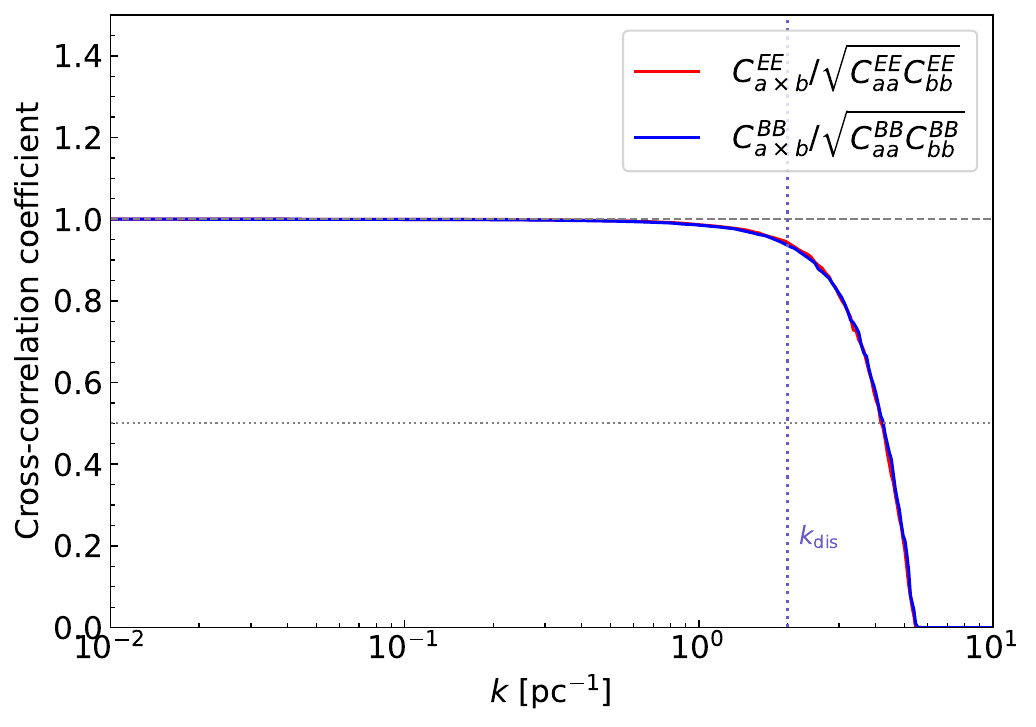}
\caption{Cross-spectrum for two independent sets of Stokes maps $a$ and $b$ from CNM using $2048^3$ resolution. $k_{\rm dis}$ is the numerical dissipation wavenumber. The dashed and dotted black lines represent the cross-correlation coefficient $r=1$ and $r=0.5$, respectively.}
\label{fig:cs}
\end{figure*}

The CNM occupies a small volume-filling fraction of the simulation domain, which may result in sparse three-dimensional sampling. When projected along the LOS to produce 2D Stokes maps, this sparsity introduces a shot-noise contribution to the angular power spectra that can masquerade as a physical signal. To quantify the scale at which this noise begins to dominate, we employ a random cell-split cross-spectrum test.

For each cell identified as CNM, we draw an independent Bernoulli random variable with equal probability and assign the cell to one of two groups. Both groups sample the identical physical volume---and therefore share the same large-scale density and magnetic field structures---but their shot-noise realizations are statistically independent. We then project each group separately into two independent sets of Stokes maps $a$ and $b$. We decompose each set of maps into $E$- and $B$-mode polarization and define the cross-correlation coefficient
\begin{equation}
    r^{XX}(k) = \frac{C_{a\times b}^{XX}(k)}
                     {\sqrt{C_{aa}^{XX}(k)\;C_{bb}^{XX}(k)}},
\end{equation}
where $X \in \{E, B\}$. At wavenumbers where the physical signal fully dominates, $r \approx 1$, while $r=0.5$ means signal-to-noise ratio = 1; at scales where sparsity noise fully dominates the power budget, $r \sim 0$. This diagnostic therefore identifies the wavenumber above which the CNM auto-spectra are unreliable (i.e., $r<0.5$) due to shot-noise contamination. Fig.~\ref{fig:cs} shows that $r \approx 1$ within the inertial range ($0.02\lesssim k \lesssim 2\,\mathrm{pc}^{-1}$), suggesting the CNM-based polarization maps are not dominated by shot noise. 

\bibliography{sample631}{}
\bibliographystyle{aasjournal}



\end{document}